\documentclass[aps,prb,twocolumn,showpacs,groupedaddress]{revtex4-1}
\usepackage{amssymb}

\usepackage{graphicx}

\begin{document}

\title[]{Magnetic field-induced spontaneous polarization reversal in multiferroic Mn$_{0.85}$Co$_{0.15}$WO$_4$}

\author{N. Poudel$^1$, K.-C. Liang$^1$, Y.-Q. Wang$^1$, Y. Y. Sun$^1$, B. Lorenz$^1$, F. Ye$^2$, J. A. Fernandez-Baca$^{2,3}$, and C. W. Chu$^{1,4}$}

\affiliation{$^1$ TCSUH and Department of Physics, University of Houston, Houston, TX 77204, USA}
\affiliation{$^2$ Quantum Condensed Matter Division, Oak Ridge National Laboratory, Oak Ridge, TN 37831-6393, USA}
\affiliation{$^3$ Department of Physics and Astronomy, The University of Tennessee, Knoxville, TN 37996-1200, USA}
\affiliation{$^4$ Lawrence Berkeley National Laboratory, 1 Cyclotron Road, Berkeley, CA 94720, USA}

\begin{abstract}
The magnetic and ferroelectric properties of the multiferroic system Mn$_{1-x}$Co$_x$WO$_4$ (x=0.135, 0.15, and 0.17) are studied in magnetic fields $H_c$ oriented along the monoclinic $c$-axis. Mn$_{0.85}$Co$_{0.15}$WO$_4$, which is right at the phase boundary between two helical spin structures, exhibits a spontaneous sign change of the ferroelectric polarization when cooled in fields $H_c>$ 25 kOe. The origin of the ferroelectric polarization is studied and two magnetic exchange interactions contributing to the polarization are identified. In Mn$_{0.85}$Co$_{0.15}$WO$_4$ domains of the characteristic helical spin structures, known for x$<$0.15 and x$>$0.15, coexist and form domain boundaries. The contributions of the different domains to the global polarization are determined. The polarization reversal in Mn$_{0.85}$Co$_{0.15}$WO$_4$ can be explained by a combination of various contributions to the polarization and a strong correlation between magnetic domains of different helical spin orders resulting in a smooth transition across the domain walls which preserves the chirality of the spin spiral.
\end{abstract}


\maketitle

\section{Introduction}
Multiferroic and magnetoelectric materials have attracted renewed attention since it was shown experimentally\cite{kimura:03} as well as theoretically\cite{mostovoy:06} that highly frustrated magnetic orders, breaking the spatial inversion symmetry, can give rise to a ferroelectric state with a macroscopic polarization which originates either from the displacement of the electronic charge distribution\cite{katsura:05} or from a polar displacement of ionic charges once the spin-lattice coupling is sufficiently strong.\cite{sergienko:06} The coupling between magnetic and ferroelectric orders within one and the same bulk phase has particularly inspired researchers studying the fundamental properties of multiferroics and their change in external magnetic and electric fields as well as the possibilities of developing future applications as memory elements, magnetoelectric sensors, etc. based on these materials.\cite{fiebig:05,spaldin:05,tokura:07} The frustrated nature of the magnetic state in most multiferroics is the reason for the extreme sensitivity of the ferroelectric/magnetic orders to small perturbations in form of magnetic and electric fields,\cite{higashiyama:04,hur:04,taniguchi:06,seki:08,sagayama:08} external pressure,\cite{delacruz:07,chaudhury:07,chaudhury:08,delacruz:08b} and chemical substitutions.\cite{seki:07,kanetsuki:07}

Among different frustrated magnetic structures giving rise to a multiferroic state, the transverse helical spin order has been identified as one of the more prominent candidates to stabilize a ferroelectric state in different compounds, as for example in TbMnO$_3$,\cite{kenzelmann:05} Ni$_3$V$_2$O$_8$,\cite{lawes:05} LiCuVO$_4$,\cite{yasui:08} CoCr$_2$O$_4$,\cite{yamasaki:06} MnWO$_4$,\cite{arkenbout:06,taniguchi:06} and many others. MnWO$_4$, also known as the mineral H\"{u}bnerite, is one end member of the wolframite family Mn$_{1-x}$Fe$_x$WO$_4$, which was known already 500 years ago.\cite{agricola:1546} However, the multiferroic properties of MnWO$_4$ have been discovered only in 2006.

The ferroelectricity observed in MnWO$_4$ between $T_C$=12.7 K and $T_L$=7.5 K is induced by an incommensurate (ICM) inversion symmetry breaking helical magnetic order (AF2 phase) which is sandwiched between an ICM sinusoidal phase (AF3 phase, $T_C<T<T_N$ with $T_N$=13.6 K) and a commensurate (CM) phase with the frustrated $\uparrow\uparrow\downarrow\downarrow$ spin modulation along the monoclinic $a$- and $c$-axes (AF1 phase, $T<T_L$).\cite{lautenschlager:93,arkenbout:06,taniguchi:06} The AF3 and AF1 phases are both collinear and paraelectric since their magnetic structures preserve the inversion symmetry. The magnetic exchange coupling and anisotropy parameters of MnWO$_4$ can be well controlled by chemical substitutions of other nonmagnetic (Zn) or magnetic transition metals (Fe, Co, Ni, Cu) since all TMnO$_4$ (T=Fe, Co, Ni, Zn) form isostructural compounds with similar lattice parameters as MnWO$_4$,\cite{weitzel:70,kleykamp:80,takagi:81} only CuWO$_4$ crystallizes in a different, slightly distorted P1 structure.\cite{arora:88} The modification of the microscopic magnetic parameters through ionic substitutions allows for the control of various magnetic structures as well as the tuning of the multiferroic properties leading to a deeper understanding of the complex physics of the multiferroic state.

While the effect of substituting nonmagnetic ions like Zn or Mg for Mn did stabilize the multiferroic AF2 phase as the ground state by suppressing the AF1 phase,\cite{meddar:09,chaudhury:11} the replacement of Mn by Fe resulted in a quick suppression of the AF2 phase and only two paraelectric phases, AF3 and AF1, survived Fe doping levels above 5\%.\cite{garciamatres:03,chaudhury:08,chaudhury:09b} The competition of the AF1 and AF2 phases for the ground state and the opposite effects of Zn/Mg and Fe substitution was attributed to the strong magnetic frustration due to competing exchange interactions and the strength of the single spin anisotropy controlled by the substitution of different ions,\cite{chaudhury:08} in agreement with a more quantitative theoretical treatment based on a Landau theory for the Heisenberg model with single-ion anisotropy.\cite{matityahu:12}

\begin{figure}
\begin{center}
\includegraphics[angle=0, width=3.3 in]{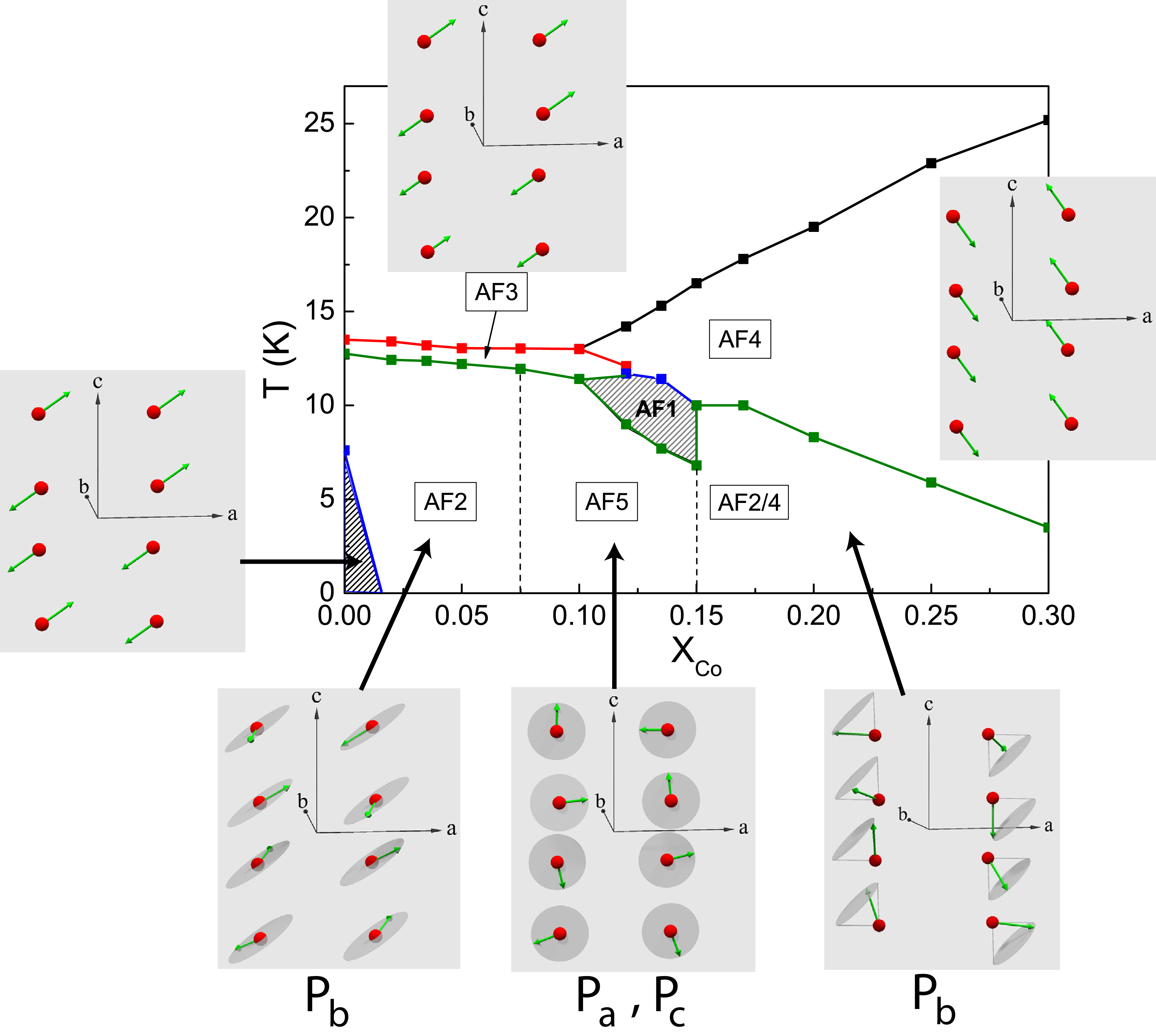}
\end{center}
\caption{(Color online) Multiferroic phase diagram of Mn$_{1-x}$Co$_x$WO$_4$ in the absence of magnetic fields.\cite{liang:12c} The different collinear and noncollinear magnetic orders in all magnetic phases are schematically shown and assigned to separate regions of the phase diagram. The two regions with line patterns have the same characteristic AF1 magnetic modulation ($\uparrow\uparrow\downarrow\downarrow$), however, the spins in the high-temperature AF1 phase (0.1$<$x$<$0.15) form an angle of -33$^\circ$ with the $a$-axis.\cite{ye:12}}
\label{Fig00}
\end{figure}

The spins of the magnetic transition metal ions Fe, Ni, and Cu in the respective structures of FeWO$_4$, NiWO$_4$, and CuWO$_4$ show uniaxial anisotropic properties very similar to the Mn ion in MnWO$_4$. Their spin easy axes lie in the $a-c$ plane at a positive angle with the $a$-axis between 57$^\circ$ and 28$^\circ$.\cite{weitzel:77} However, the magnetic anisotropy in CoWO$_4$ is very different from the other transition metals; the Co spins form an angle of -46$^\circ$ with the $a$-axis,\cite{forsyth:94} nearly perpendicular to the Mn spins in MnWO$_4$. As a consequence, the effects of Co doping on the magnetic phases and on the nature of the spiral spin order in Mn$_{1-x}$Co$_x$WO$_4$ are more complex than in other cases.

\begin{figure}
\begin{center}
\includegraphics[angle=0, width=3 in]{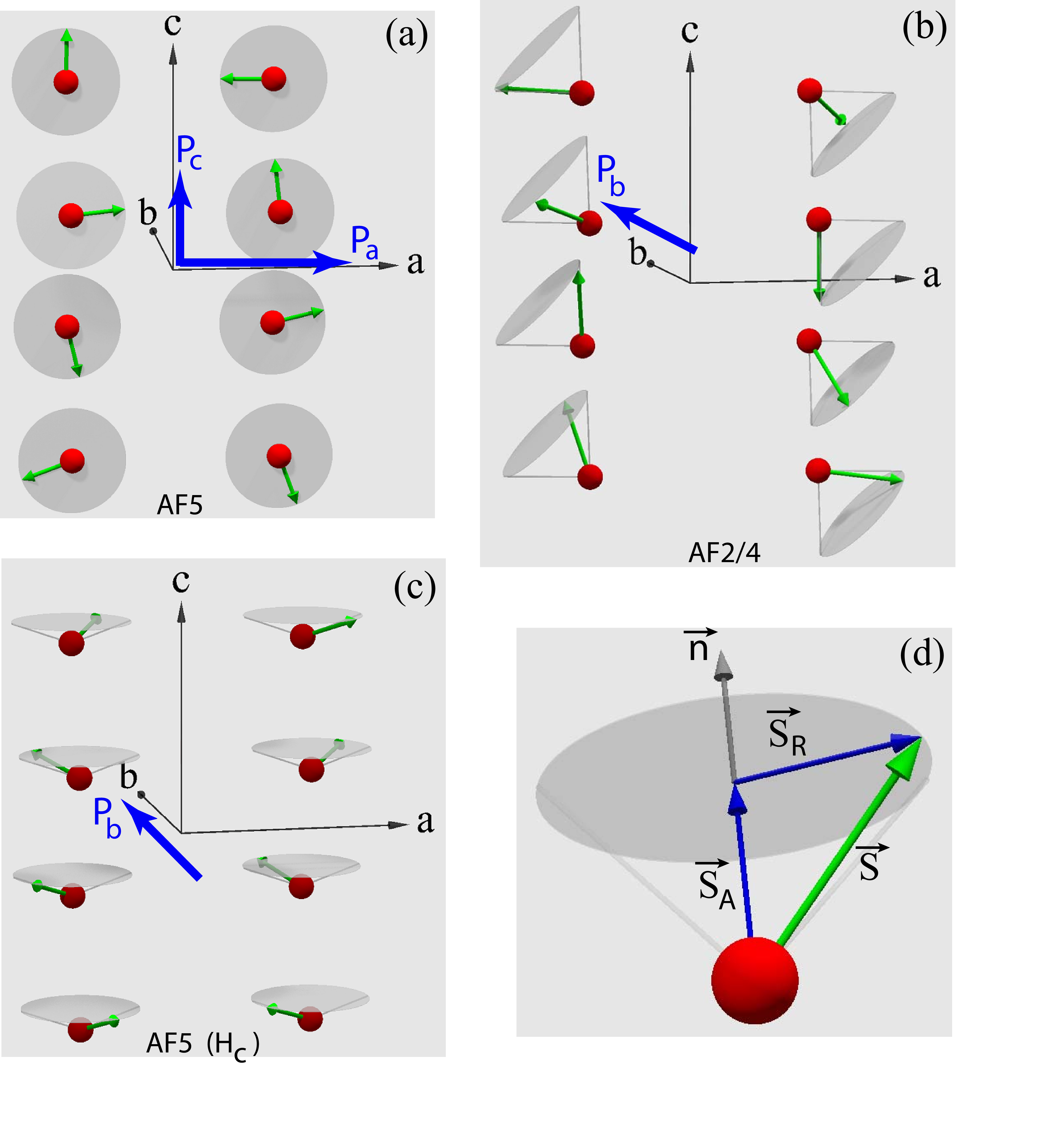}
\end{center}
\caption{(Color online) Schematic presentation of different spin orders in the multiferroic phases which coexist in Mn$_{0.85}$Co$_{0.15}$WO$_4$, as discussed in the text. (a) $a-c$ spiral spin structure of the AF5 phase, stable for 0.075$<$x$<$0.15. (b) Conical spin structure of the AF2/4 phase, stable for x$>$0.15. (c) Spin structure of the AF5 phase after the spin-flop in $c$-axis magnetic fields. (d) Axial ($\vec{S}_A$) and radial ($\vec{S}_R$) components of the spin vector $\vec{S}$ in the conical phase. The normal vector $\vec{n}$ defining the orientation of the spiral is also shown. The bold (blue) arrows in (a) to (c) show the components of the ferroelectric polarization allowed by symmetry when magnetic exchange interactions along $c$- and $a$-axes are taken into account.}
\label{Fig0}
\end{figure}

A completely different and more complex phase diagram was recently revealed for the case of Co substitution.\cite{song:09,liang:12c,ye:12} Besides the suppression of the AF1 phase by a small amount of Co (similar to Zn doping), two more sudden changes of the magnetic order in the multiferroic phase upon increasing Co content have been reported.\cite{song:10,olabarria:12,olabarria:12b,ye:12} At about 7.5\% Co content the spin helix in the AF2 phase suddenly rotates by 90$^\circ$ into the $a-c$ plane causing a flop of the ferroelectric polarization from the $b$- to the $a$-axis (AF5 phase). At 15\% Co doping, the $a-c$ spiral becomes unstable and the spins form a conical structure about the easy spin axis (AF2/4 phase). This conical spin modulation is characterized by two $\overrightarrow{Q}$ vectors, one CM modulation $\overrightarrow{Q}_4$=(0.5,0,0) describing the modulation of the axial component $S_A$ and a the second ICM modulation $\overrightarrow{Q}_2$=(-0.211,0.5,0.452) defining the radial component $S_R$ of the spin vector $\overrightarrow{S}$. The complete phase diagram of Mn$_{1-x}$Co$_x$WO$_4$ and the spin structures of the different collinear and noncollinear magnetic phases are shown in Fig. \ref{Fig00}. At 15\% Co content two multiferroic phases (AF5 and AF2/4) coexist at low temperatures.\cite{chaudhury:10} The spin orders in the two coexisting phases are schematically shown in Fig. \ref{Fig0}a and Fig. \ref{Fig0}b, respectively. Fig. \ref{Fig0}d defines the axial and radial components of the spin in the conical structure. Similar to the AF2 phase, the conical phase allows only for a $b$-axis component of the ferroelectric polarization, i.e. Mn$_{1-x}$Co$_x$WO$_4$ shows another polarization flop from $a$ to $b$ at x=0.15.\cite{liang:12c} The symmetry-allowed components of the polarization are shown by bold (blue) arrows in Fig. \ref{Fig0}.

With this multitude of different phases in the phase diagram of Mn$_{1-x}$Co$_x$WO$_4$, some multiferroic but others collinear, highly frustrated and paraelectric, the influence of external perturbations such as magnetic fields can be significant and interesting, particularly near the critical Co concentrations where the multiferroic states and the direction of the ferroelectric polarization $\overrightarrow{P}$ change abruptly (x$_{c1}\simeq$0.05 and x$_{c2}\simeq$0.15). In the low-doping range (x=0.05) it was shown that the direction of $\overrightarrow{P}$ rotated continuously from the $b$-axis toward the $a$-axis upon increasing magnetic fields oriented along the $b$-axis.\cite{liang:12} In the AF5 phase (for x=0.1) it was found that the $a$-axis polarization was strongly suppressed in a $c$-axis field.\cite{song:10,olabarria:12} A systematic study of the field effects on the magnetic and multiferroic phases of Mn$_{1-x}$Co$_x$WO$_4$ for x$>$0.1 is still lacking.

Here we study the magnetic field effects on the multiferroic phases in Mn$_{1-x}$Co$_{x}$WO$_4$ for x=0.135, 0.15, and 0.17, three compositions that are crossing the boundary between the $a-c$ spiral (AF5) and the conical (AF2/4) phases. For x=0.15, we find a spontaneous polarization reversal at a critical temperature in a constant external field applied along the $c$-axis. The results are interpreted in terms of two coexisting multiferroic phases with a strong correlation across the walls separating the domains of these phases.

\section{Experimental}
Single crystals of Mn$_{1-x}$Co$_{x}$WO$_4$ (x=0.135, 0.15, 0.17) have been grown in a floating zone optical furnace, as described earlier.\cite{chaudhury:10,liang:12c} Powder X-ray analysis confirms the monoclinic structure (space group P2/c) and no impurity phases could be detected. The precise chemical composition was determined by testing different spots of one single crystal employing inductively coupled plasma mass spectrometry. The measured Co content was always close to the nominal composition with a high uniformity across the crystal. The lattice parameters determined from powder X-ray spectroscopy show a tendency to decrease only slightly with the Co doping from those of MnWO$_4$. The crystals cleave easily and expose a smooth $a-c$ plane after cleavage. The out-of-plane and in-plane orientations were determined by single crystal Laue diffraction. Crystals for different measurements were cut from the bigger piece and shaped according to the demands.

Magnetic measurements have been conducted using the Magnetic Property Measurement System (MPMS, Quantum Design) in magnetic fields up to 50 kOe. The ferroelectric polarization in the multiferroic state was determined by measuring the pyroelectric current. A thin (typical 0.5 mm) plate-like crystal was prepared for the measurements. Silver paint was used to form a contact area of about 20 to 40 mm$^2$. Temperature and magnetic field control was provided by the Physical Property Measurement System (PPMS, Quantum Design) and a home made top-loading dielectric probe was employed as sample holder. The alignment of ferroelectric domains was secured through cooling the sample from above T$_N$ in an external electric bias field of about 2 to 3 kV/cm. At the lowest temperature, the bias field was released and the two contacts were shorted for several minutes. The pyroelectric current, measuring the change of the ferroelectric polarization, was recorded upon heating at a speed of 1 K/min employing a K6517A electrometer (Keithley). The polarization was calculated by integrating the current starting from high temperature (paraelectric state).

\section{Results and Discussions}
\subsection{Polarization reversal for x=0.15}
\label{P15}
\begin{figure}
\begin{center}
\includegraphics[angle=0, width=2.5in]{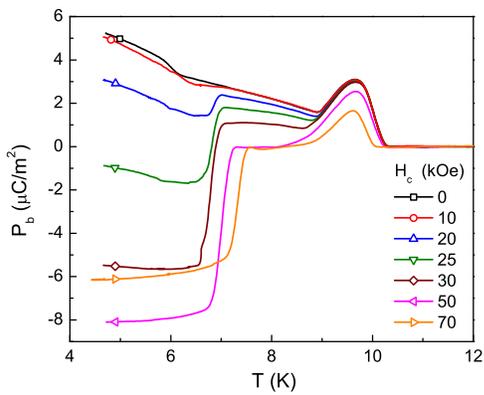}
\end{center}
\caption{(Color online) Ferroelectric polarization of Mn$_{1-x}$Co$_{x}$WO$_4$ (x=0.15) in magnetic fields oriented along the $c$-axis. Data shown were measured with increasing temperature.}
\label{Fig1}
\end{figure}

In the absence of magnetic fields the ferroelectric polarization in the multiferroic phase is directed along the $b$-axis.\cite{chaudhury:10} The characteristic feature of $P_b(T)$ is a peak just below the transition temperature $T_{C1}$=10.2 K followed by a continuous increase upon further decreasing $T$, as shown in Fig. \ref{Fig1}. Neutron scattering experiments have shown that this $b$-axis polarization originates from the formation of the conical AF2/4 phase which coexists with other magnetic phases. The scattering intensity associated with the AF2/4 phase has a similar sharp peak just below $T_{C1}$ and thus mimics the temperature dependence of $P_b$.\cite{chaudhury:10,ye:12} The decrease of the polarization on the low-temperature side of the peak is due to the formation of another collinear magnetic phase with the same modulation as the AF1 phase of the undoped compound ($\overrightarrow{Q}_{AF1}=(\pm0.25,0.5,0.5)$) which is stable between 6.5 K and 10 K. This phase, that was observed in Mn$_{1-x}$Co$_x$WO$_4$ for 0.12$\leq$x$\leq$0.15,\cite{ye:12} is paraelectric and it competes with the ferroelectric AF2/4 phase. It should be noted that below 6.5 K the AF5 phase ($a-c$ spiral) coexists with the AF2/4 phase, according to the neutron scattering data. While the AF5 phase is responsible for the large $a$-axis polarization at x=0.1, it was shown that the magnitude of P$_a$ quickly decreases with increasing x and it becomes vanishingly small at x=0.15.\cite{liang:12c}

The effect of an external $c$-axis magnetic field on the ferroelectric polarization is shown in Fig. \ref{Fig1}. At low fields ($H_c<$ 20 kOe) $P_b$(T) initially does not change significantly. At about 20 kOe $P_b$ develops a sharp, step-like decrease near $T_{C2}\simeq$7 K with decreasing temperature, indicating a transition in the magnetic system that affects the ferroelectric polarization. With further increasing $H_c$ the step size increases and $P_b$ reverses sign below $T_{C2}$ for fields above 25 kOe. It is most interesting that the sign change of $P_b$ is even observed during the cooling process with a positive poling voltage applied. Measurements of the pyroelectric current while cooling show that the transition into the low-temperature phase happens about 2 K lower indicating a first order transition at $T_{C2}$ with a significant thermal hysteresis (note that only data obtained with increasing temperature are displayed in Fig. \ref{Fig1} for better clarity). The critical temperature $T_{C2}$ increases by about 0.5 K between 20 kOe and 70 kOe while $T_{C1}$ decreases slightly between 50 kOe and 70 kOe.

It should be noted that a polarization reversal was recently observed in the undoped MnWO$_4$ upon increasing magnetic field, but only at very high magnetic fields (H$>$140 kOe).\cite{mitamura:12} The change from the AF2 spin spiral to a conical spin order with a different orientation of the axial spin vector induced by high magnetic fields was suggested as a possible explanation and a theoretical description based on a Landau theory could well reproduce the experimental high-field phase diagram of MnWO$_4$.\cite{quirion:13} However, the polarization reversal in MnWO$_4$ was only observed upon changing the magnetic field at a constant temperature and the derived phase diagrams, experimental as well as theoretical, do not indicate the possibility of a sign reversal as a function of temperature. In contrast, the sign change of $P_b$ in Mn$_{0.85}$Co$_{0.15}$WO$_4$ could be observed as a function of temperature in constant magnetic fields, as shown in Fig. \ref{Fig1}. This implies that different physical mechanisms should be responsible for the polarization reversal in Mn$_{0.85}$Co$_{0.15}$WO$_4$.

In order to understand the origin of the sign reversal of $P_b$ in fields above 25 kOe we have to consider all magnetic phases that could contribute to $P_b$. Mn$_{0.85}$Co$_{0.15}$WO$_4$ is right at the boundary between the AF5 ($a-c$ spiral) and AF2/4 (conical) phases and it is not surprising that both multiferroic phases have been found to coexist at low temperatures.\cite{ye:12} The conical spin structure is consistent with a $b$-axis polarization whereas the $a-c$ spiral can generate polarizations within the $a-c$ plane, but no component along $b$. This was indeed observed experimentally\cite{liang:12c} and it is consistent with general symmetry arguments.\cite{mostovoy:06} The ferroelectric polarization of the AF5 magnetic phase, which has a maximum between x=0.075 and x=0.1, does decrease for x$>$0.1 and it did vanish at x=0.15.\cite{liang:12c} However, this may change under the action of an external magnetic field. It is therefore necessary to study the ferroelectric polarization of the AF5 and AF2/4 phases in magnetic fields separately. Neutron scattering experiments have revealed that for slightly smaller (larger) Co concentrations, i.e. x=0.135 (0.17), the low temperature phase is the pure AF5 (AF2/4) phase.\cite{ye:12} These two special Co concentrations are used to investigate the $c$-axis field effect on $P_b$ in both phases independently.

\subsection{Field effects on the polarization for x=0.135 and x=0.17}
\label{P2}
\begin{figure}
\begin{center}
\includegraphics[angle=0, width=2.5 in]{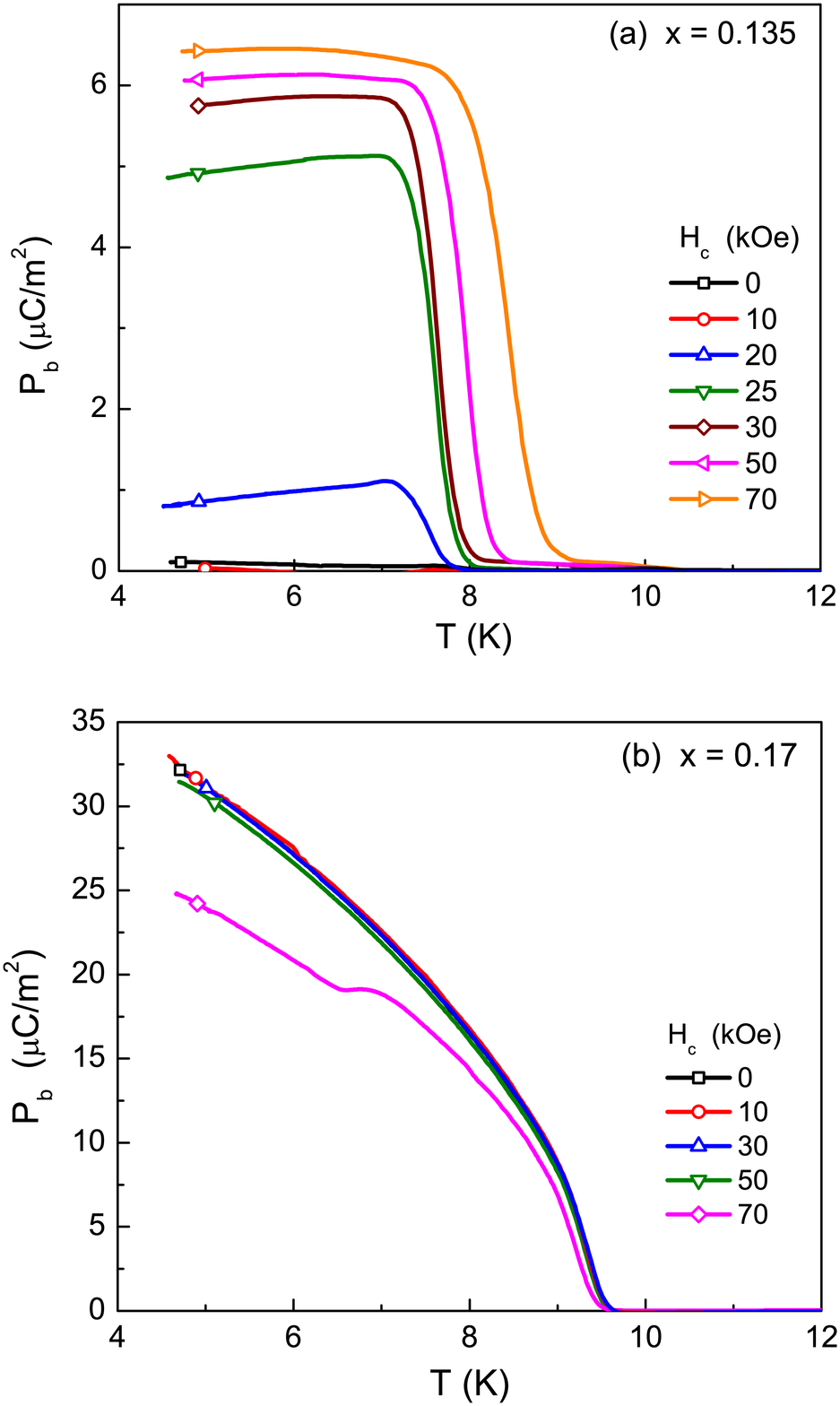}
\end{center}
\caption{(Color online) Ferroelectric polarization of Mn$_{1-x}$Co$_{x}$WO$_4$ in magnetic fields oriented along the $c$-axis. (a): x=0.135, (b): x=0.17. Data shown were measured with increasing temperature.}
\label{Fig2}
\end{figure}

The $b$-axis polarization for x=0.135 (AF5 phase) in $c$-axis magnetic field is shown in Fig. \ref{Fig2}a. At low fields, $P_b$ is zero within the experimental resolution. It should be noted that even the $a$-axis component of $\overrightarrow{P}$ is very small ($P_a<$2 $\mu$C/m$^2$).\cite{liang:12c} However, with increasing $H_c$ above 20 kOe there arises a sizable polarization along the $b$-axis at temperatures below 8 K. This polarization component increases further with increasing field and the critical temperature also increases to about 9 K at 70 kOe. The stabilization of $P_b$ in a $c$-axis magnetic field has to be a consequence of a spin-flop transition from the $a-c$ spiral to an umbrella-like magnetic structure, as schematically shown in Figs. \ref{Fig0}a and c. The umbrella structure represents a conical spin modulation with the axial component $S_A$ aligned with the field, i.e. along the $c$-axis.

A magnetic field-induced spin-flop transition is frequently observed in antiferromagnetic spin structures when the magnetic field is parallel to the spin orientation. Above a critical field, the spins reorient themselves perpendicular to the field, conserving the antiferromagnetic exchange energy. The simplest example to describe this effect is the Ising model in a transverse field.\cite{stinchcombe:73} For the more complex case of non-collinear spin structures, the spin-flop transition from a proper screw to a conical modulation was theoretically studied\cite{nagamiya:62,kitano:64} and it was indeed experimentally observed in Mn$_{1-x}$Co$_x$WO$_4$ for x=0.1.\cite{olabarria:12} In section \ref{mag} we provide further evidence for the spin-flop from magnetic measurements. The origin of $P_b$ in the spin-flop phase will be discussed in detail in section \ref{origin}.

The $b$-axis polarization in the AF2/4 conical phase at x=0.17 responds very differently to the $c$-axis magnetic field, as shown in Fig. \ref{Fig2}b. $P_b$ is not affected at all in fields up to 50 kOe. Only at higher fields the maximum of $P_b$ decreases slightly from 32 $\mu$C/m$^2$ at 50 kOe to 25 $\mu$C/m$^2$ at 70 kOe. The ferroelectric critical temperature $T_C$ decreases by about 0.1 K. Alternative magnetic field orientations along the $a$- or $b$-axes have shown a similar result, a small depletion of the maximum $P_b$ and a minute decrease of $T_C$. It should be noted that no component of $\overrightarrow{P}$ along other directions ($a$ or $c$) have been observed in any field up to 70 kOe. The conical AF2/4 magnetic structure, that defines the ground state for x$>$0.15, appears to be very stable with respect to magnetic fields up to 70 kOe.

From the results presented in Fig. \ref{Fig2} it becomes clear that a $b$-axis ferroelectric polarization can originate from both phases, AF2/4 as well as AF5, in the latter phase a $c$-axis magnetic field above 20 kOe needs to be present and the magnetic system has to be in the spin-flop phase. Since the AF2/4 and AF5 phases coexist in Mn$_{1-x}$Co$_x$WO$_4$ for x=0.15, one should expect two contributions to $P_b$ arising at different temperatures, $T_{C1}$ and $T_{C2}$. It remains to be shown that the spin-flop actually happens in Mn$_{0.85}$Co$_{0.15}$WO$_4$.

\subsection{Evidence for the spin-flop in Mn$_{0.85}$Co$_{0.15}$WO$_4$}
\label{mag}
Magnetic systems with strong uniaxial spin anisotropy, such as Mn$_{1-x}$Co$_x$WO$_4$, display a characteristic response of the temperature dependent magnetization to the external magnetic field below the antiferromagnetic transition temperature $T_N$. If the field is oriented longitudinal (parallel) to the spin moment (spin easy axis), the magnetization $M(T)$ drops sharply below $T_N$ and continues to decrease toward lower temperatures. However, in the transverse field case $M(T)$ remains nearly constant below $T_N$.\cite{stinchcombe:73} The same behavior is also observed in spiral spin structures. Once the external field is applied in the helical plane the spin spiral has a longitudinal component relative to the field and $M(T)$ decreases below $T_N$. However, the magnetization below $T_N$ is nearly independent of temperature once the field is oriented perpendicular to the helical plane. Antiferromagnetically ordered spins with a component longitudinal to the field tend to experience a spin flop above a critical field which reorients the spins so that they become nearly perpendicular to the field. This spin flop is associated with a change of the characteristic $M(T)$ behavior below $T_N$ as discussed above. Similarly, the application of a magnetic field in the plane of a spin spiral will flop the spins forming a conical, umbrella-like structure with the corresponding change in the temperature dependence of $M(T)$.

\begin{figure}
\begin{center}
\includegraphics[angle=0, width=2.5 in]{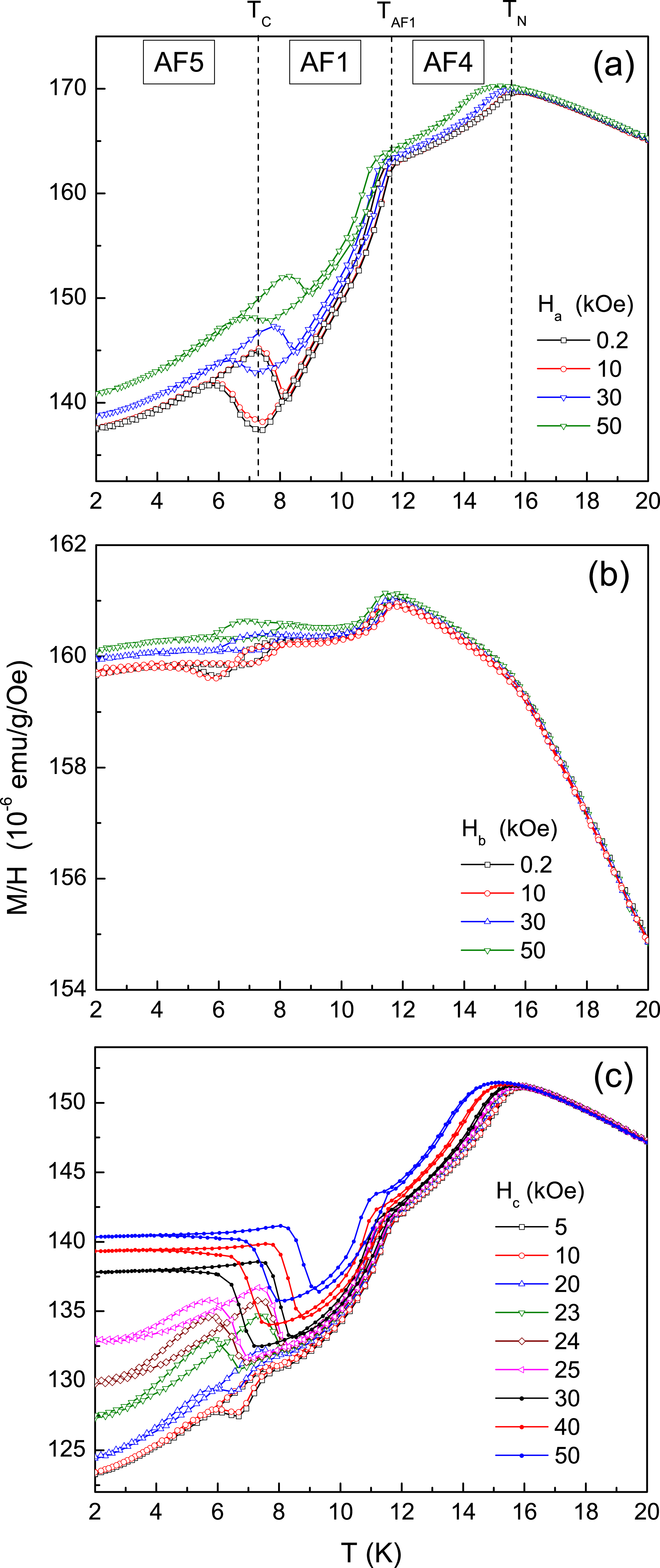}
\end{center}
\caption{(Color online) Magnetization M/H of Mn$_{1-x}$Co$_x$WO$_4$ (x=0.135) in fields of different orientations up to 50 kOe. The three magnetic phases are separated by dashed lines in (a) and labeled accordingly. The transition temperatures into the AF4 phase ($T_N$), the AF1 phase ($T_{AF1}$), and the AF5 phase ($T_C$) are also indicated at the top of the graph. Here and in Figs. \ref{Fig4} and \ref{Fig6} the low-field data are slightly scaled up to match the high-temperature data for a better comparison of the various phase transitions.}
\label{Fig3}
\end{figure}

The magnetic data for x=0.135 in Fig. \ref{Fig3}a confirm the field dependence of $M(T)$ as discussed above. At zero magnetic field the magnetic system passes through three transitions with decreasing temperature.\cite{liang:12c,ye:12} The onset of commensurate and collinear magnetic order (described by the AF4 magnetic structure) is clearly marked by a slope change of $M(T)$ at $T_N$. This magnetic structure is replaced by the frustrated $\uparrow\uparrow\downarrow\downarrow$ collinear AF1 phase below $T_{AF1}$, also clearly defined by another change of slope. At about 7 K, the spiral magnetic structure with the helical plane in the $a-c$ plane becomes the stable magnetic structure (AF5 phase). The transition into the AF5 phase shows a hysteretic anomaly in $M(T)$. Since in the AF5 magnetic structure the spins form a spiral confined to the $a-c$ plane, we would expect a spin-flop transition in fields along the $a$- and/or $c$-axes.

The magnetic response to an $a$-axis field, shown in Fig. \ref{Fig3}a, does not seem to indicate a spin-flop within the field limits of 50 kOe. However, above 30 kOe there is a sizable increase of $M(T)$ which could be the beginning of the spin reorientation. Larger fields have to be applied to induce the expected flop of the magnetic moments. This is consistent with data published for Mn$_{0.9}$Co$_{0.1}$WO$_4$ showing the onset of the spin-flop in $a$-axis fields above 60 kOe.\cite{olabarria:12} Magnetic fields along the $b$-axis are already perpendicular to the spins of the spiral phase. Therefore, the $M/H$ data in Fig. \ref{Fig3}b show little dependence on the field at low temperatures. With the field applied along the $c$-axis, however, the magnetization below $T_C$ experiences a sharp increase in a narrow field range between 20 and 30 kOe. At higher fields $M(T)$ is nearly constant below $T_C$ as expected after the spins flop perpendicular to the field. The resulting spin spiral is now in the $a-b$ plane with a small conical component induced by the field (see Fig. \ref{Fig0}c) and it gives rise to a $b$-axis ferroelectric polarization, as discussed in more detail in section \ref{origin}. The critical temperatures $T_C$ derived from the magnetic data (upon heating) are in perfect agreement with the onset of $P_b$ in Fig. \ref{Fig2}a.

\begin{figure}
\begin{center}
\includegraphics[angle=0, width=2.5 in]{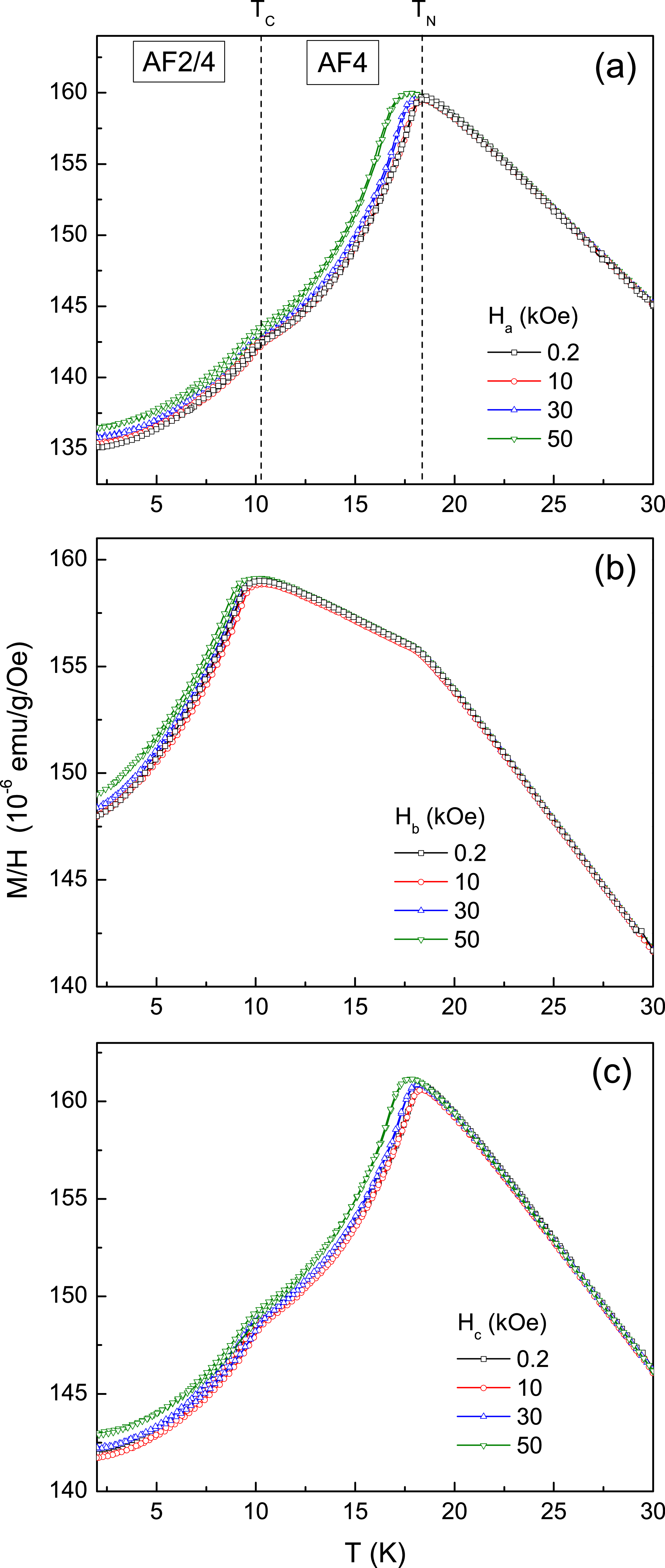}
\end{center}
\caption{(Color online) Magnetization M/H of Mn$_{1-x}$Co$_x$WO$_4$ (x=0.17) in fields of different orientations up to 50 kOe. The two magnetic phases are separated by dashed lines in (a) and labeled accordingly. The transition temperatures into the AF4 phase ($T_N$) and the AF2/4 phase ($T_C$) are also indicated at the top of the graph.}
\label{Fig4}
\end{figure}

The magnetization response to high fields is very different in the conical AF2/4 phase for x=0.17. Fig. \ref{Fig4} shows $M/H$ data up to 50 kOe along all three crystallographic orientations. A sharp drop of the magnetization below $T_N$ in fields along the $a$- and $c$-axes (Figs. \ref{Fig4}a and \ref{Fig4}c) is consistent with a longitudinal component of $M$ in the AF4 phase. Neutron scattering experiments have determined the spin structure of the AF4 phase as collinear with a CM magnetic structure defined by $\overrightarrow{Q}_4$ with a strong uniaxial anisotropy. The spin-easy axis is in the $a-c$ plane at an angle of about -50$^\circ$ with respect to the $a$-axis.\cite{olabarria:12,ye:12} The $M/H$ characteristics between $T_C$ and $T_N$ in a $b$-axis field, shown in Fig. \ref{Fig4}b, is quite different with $M/H$ still increasing with decreasing temperature due to the perpendicular orientation of the spin with respect to $H_b$.

In the conical AF2/4 phase (see Fig. \ref{Fig0}b) the spin has longitudinal components along all three field orientations, $a$, $b$, and $c$. Therefore, $M/H$ is decreasing with temperature below $T_C$. It is remarkable that $M/H$ in Fig. \ref{Fig4} shows very little change in magnetic fields up to 50 kOe. This is consistent with the small field effect on the polarization discussed above (and shown in Fig. \ref{Fig2}b) for x=0.17. The conical AF2/4 phase is most stable with respect to external magnetic fields up to 50 kOe.

\begin{figure}
\begin{center}
\includegraphics[angle=0, width=2.5 in]{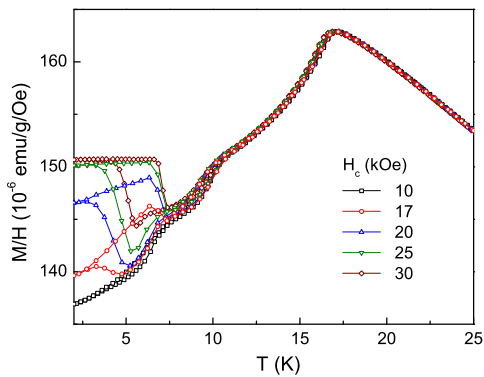}
\end{center}
\caption{(Color online) Magnetization M/H of Mn$_{0.85}$Co$_{0.15}$WO$_4$ in magnetic fields oriented along the $c$-axis.}
\label{Fig6}
\end{figure}

The magnetization data for x=0.15 are shown in Fig. \ref{Fig6} with the magnetic field applied along the $c$-axis. The data clearly reveal the occurrence of a spin-flop transition in Mn$_{0.85}$Co$_{0.15}$WO$_4$, similar to the data in Fig. \ref{Fig3}c for x=0.135. It is therefore conceivable that the spin-flop transition in Mn$_{0.85}$Co$_{0.15}$WO$_4$ arises from the AF5 $a-c$ spiral phase that coexists with the AF2/4 conical phase. The critical field of the spin-flop transition as derived from the magnetic data of Fig. \ref{Fig6} coincides with the critical field above which the polarization reversal (Fig. \ref{Fig1}) had been observed. In addition, the upward shift of the critical temperature $T_{C2}$ of the polarization reversal with increasing field $H_c$ is consistent with the similar temperature shift of the spin-flop transition (Fig. \ref{Fig6}). This leads to the conclusion that the polarization reversal observed in $c$-axis magnetic fields in Mn$_{0.85}$Co$_{0.15}$WO$_4$ is associated with the spin-flop of the magnetic structure in the AF5 phase. However, it is not clear why the induced $b$-axis polarization in the spin-flop phase is negative despite the positive poling electric field applied during the cooling process. A more detailed discussion of the possible contributions to the $b$-axis polarization is needed and presented in the next section.

\subsection{Source of the ferroelectric polarization in different phases}
\label{origin}
MnWO$_4$ and it's substituted analog Mn$_{1-x}$Co$_x$WO$_4$ exhibit a wealth of frustrated magnetic phases the origin of which lie in competing exchange interactions. Theoretical calculations\cite{tian:09} and inelastic neutron scattering measurements\cite{ehrenberg:99,ye:11} have led to a more specific picture of the exchange coupling and anisotropy parameters of MnWO$_4$. The most accurate fit of the magnetic excitations to a Heisenberg model with uniaxial anisotropy was obtained by involving 11 inequivalent exchange constants, J1 to J11, among different pairs of Mn spins.\cite{ye:11} This long-range nature of the magnetic interactions explains the extraordinary stability of the magnetic and multiferroic phases with respect to diluting the magnetic system through nonmagnetic Zn substitution of up to 50 \%.\cite{chaudhury:11} With so many relevant magnetic exchange interactions present it is important to discuss those pairs of interacting spins in the structure of MnWO$_4$ that contribute significantly to the ferroelectric polarization in the multiferroic state.

\begin{figure}
\includegraphics[angle=0, width=2.5 in]{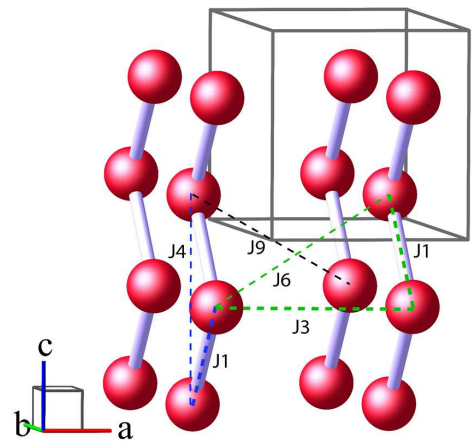}
\caption{(Color online) Mn ions of MnWO$_4$ forming zigzag chains along the $c$-axis. The most relevant exchange coupling constants are labeled Ji and the interacting ions are connected by dashed lines. Magnetically frustrated interactions are J1 and J4 (blue) as well as J1, J3, and J6 (green). Only pairs of spins coupled by J1 and J3 (bold dashed lines) contribute significantly to the ferroelectric polarization.}
\label{Fig7}
\end{figure}

In the following we will focus on the largest coupling constants which assume values within 50 \% of the maximum constant. The largest coupling constants are J1 and J6 (we use the same labels for the J's as in the original paper\cite{ye:11}) with values of about -0.42 eV. The additional significant parameters are J3 (-0.32 eV), J4 (-0.26 eV), and J9 (-0.26 eV), the corresponding exchange pathways are shown in Fig. \ref{Fig7} as dashed lines.

It is obvious that the spins in the MnO$_6$ octahedral chains aligned with the $c$-axis are frustrated since the nearest (J1) and next-nearest (J4) neighbor interactions compete with one another and the result of this competition in MnWO$_4$ is an incommensurate spin modulation with $Q_z\simeq$0.457 at high temperature and the $\uparrow\uparrow\downarrow\downarrow$ spin sequence at low $T$. The spins of neighboring chains, stacked along the $a$-axis, are also frustrated since three spins coupled by J1, J3, and J6 form a triangle with all AFM exchange interactions. This explains the high-temperature ICM modulation with $Q_x\simeq$-0.214 and the $\uparrow\uparrow\downarrow\downarrow$ spin sequence along the $a$-axis at low $T$.

It is essential to identify those pairs of spins which contribute significantly to the ferroelectric polarization in the spiral and conical states. Since the local electrical dipole moment, induced by the non-collinear spin pairs, is proportional to the vector product of the two spins, the angle between the interacting spins is important. According to the incommensurate modulations ($Q_x$, $Q_z$) and the specific helical orders as shown in Fig. \ref{Fig0}, there are only two pairs of spins which make large contributions to the electrical polarization. Those are the pairs interacting through J1 and J3 along the $c$- and $a$-directions, respectively. The spins within each pair are almost perpendicular to each other with an angle between 80$^\circ$ and 85$^\circ$, depending on the exact values of $Q_x$ and $Q_z$ in the helical phase. The vector product of those spins is therefore near its maximum value. All other spin pairs, coupled by J4, J6, and J9, are nearly collinear (parallel or antiparallel) and their contributions to the polarization can be neglected.

The two exchange pathways J1 and J3 contribute separately to the ferroelectric polarization in the different multiferroic states, noted as $\overrightarrow{P}^{(J1)}$ and $\overrightarrow{P}^{(J3)}$ hereafter. The magnitude and orientation of $\overrightarrow{P}^{(J1)}$ and $\overrightarrow{P}^{(J3)}$ depend on the specifics of the non-collinear spin structure in the multiferroic phases AF2, AF5, and AF2/4. The contributions have been derived as function of the orientation of the normal unit vector $\overrightarrow{n}$ (see Fig. \ref{Fig0}d) defining the orientation of the spin spiral in space as:\cite{liang:12c}

\begin{eqnarray}
\overrightarrow{P}^{(J1)}=C^{(J1)}m_{\parallel}m_{\perp}sin(\pi Q_z) [-sin\Theta sin\varphi \overrightarrow{e}_x\nonumber \\ +sin\Theta cos\varphi \overrightarrow{e}_y]
\end{eqnarray}

\begin{eqnarray}
\overrightarrow{P}^{(J3)}=C^{(J3)}m_{\parallel}m_{\perp}sin(2\pi Q_x)[-cos\Theta \overrightarrow{e}_y\nonumber \\ +sin\Theta sin\varphi \overrightarrow{e}_z]
\end{eqnarray}

Here $\Theta$ is the angle of $\overrightarrow{n}$ with the $z$-axis and $\varphi$ is the angle of the projection of $\overrightarrow{n}$ into the $x-y$ plane with the $x$-axis. $Q_x$ and $Q_z$ are measured in fractions of 1/$a$ and 1/$c$, respectively. The orthogonal coordinate system $x$, $y$, and $z$ is chosen so that $x\parallel a$ and $y\parallel b$ (the small one degree tilt of $c$ with respect to $z$ is neglected for simplicity). $m_\parallel$ and $m_\perp$ denote the long and short axes of the spin spiral, respectively, and $C^{(J1)}$ and $C^{(J3)}$ are constants depending on structural details and the interactions along the J1 and J3 exchange pathways. The $a-c$ spiral of the AF5 phase (Fig. \ref{Fig0}a) is defined by $\Theta=90^\circ$, $\varphi=90^\circ$ and it generates two contributions to the ferroelectric polarization, $P_a^{(J1)}$ and $P_c^{(J3)}$. Both components have been observed experimentally.\cite{song:10,olabarria:12,liang:12c} The radial components of the spin vectors in the AF2/4 phase (Fig. \ref{Fig0}b) form the spiral with $\varphi=0^\circ$ and both exchange couplings, J1 and J3, contribute to the polarization which is now oriented along the $b$-axis. In the spin-flop phase (Fig. \ref{Fig0}c), however, $\Theta$ is close to zero and the major contribution to the polarization is $P_b^{(J3)}$, resulting from the interchain exchange coupling J3 along the $a$-axis.

\subsection{Origin of the polarization flop in Mn$_{0.85}$Co$_{0.15}$WO$_4$}
\label{polarizationflop}
In Mn$_{0.85}$Co$_{0.15}$WO$_4$, two non-collinear phases coexist and may contribute to the ferroelectric polarization.\cite{ye:12} The conical AF2/4 phase emerges from the collinear AF4 phase at $T_{C1}$=10.2 K and this phase is responsible for the increase of $P_b$ below this temperature (Fig. \ref{Fig1}). $T_{C1}$ is consistent with the value of 9.6 K obtained for the onset of ferroelectricity in Mn$_{0.83}$Co$_{0.17}$WO$_4$, as shown in Fig. \ref{Fig2}b (Note that $T_{C1}$ of the AF2/4 phase decreases with increasing doping x).\cite{liang:12c,ye:12} At lower temperature and in $c$-axis magnetic fields above 20 kOe, an additional contribution to $P_b$ arises from the $a-b$ spiral of the AF5 phase in the spin flop state, as shown for x=0.135 in Fig. \ref{Fig2}a. The sudden change of $P_b$ at $T_{C2}$=7 K shown in Fig. \ref{Fig1} must therefore originate from the high-field AF5 phase (Fig. \ref{Fig0}c) which generates a $b$-axis component of the polarization (Fig. \ref{Fig2}a). The critical temperature $T_{C2}$ is also comparable with the value of 7.8 K, the critical temperature of the AF5 phase for x=0.135. Furthermore, the upward shift of $T_{C2}$ with field $H_c$ for x=0.15 is similar to the same shift for x=0.135.

\begin{figure}
\includegraphics[angle=0, width=3 in]{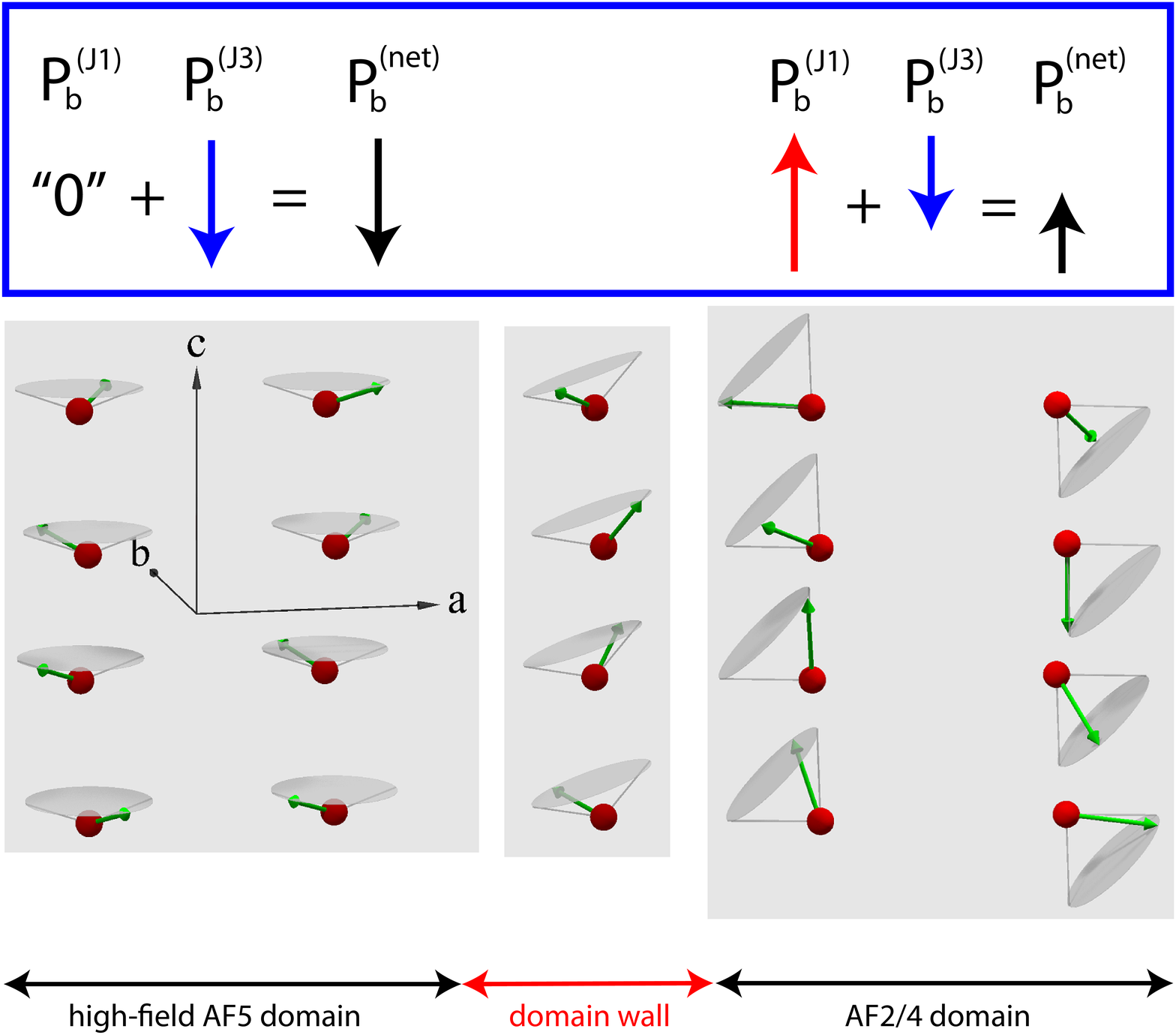}
\caption{(Color online) Possible scenario of a smooth transition from the high-field AF5 phase to the AF2/4 phase. The chirality of the helical spin modulation is maintained across the domain wall. The top box shows a schematic of how the different contributions to $P_b$ add up to the net polarization in the two domains, as discussed in the text.}
\label{Fig8}
\end{figure}

From the above discussion we conclude that two multiferroic phases contribute to the polarization $P_b$ in Mn$_{0.85}$Co$_{0.15}$WO$_4$ in high magnetic fields $H_c$. The first contribution below 10 K is from the AF2/4 conical phase and it is aligned (positive) with the poling electric field applied during the cooling process. A second contribution to $P_b$ below 7 K arises from the spin-flopped AF5 phase causing the sudden drop of $P_b$. It is most remarkable that this contribution is negative despite the positive poling voltage applied while cooling. This unusual behavior can only be explained by a strong correlation of the $b$-axis polarizations of both magnetic structures. Once $P_b^{(AF2/4)}$ of the conical AF2/4 is established and aligned with the poling field at higher temperature, the $P_b^{(AF5)}$ arising from the AF5 phase is forced into the opposite direction resulting in the sign reversal as shown in Fig. \ref{Fig1}. Since the sign of the polarization in a helical spin structure is determined by the chirality of the helix, this result requires a strong coupling of the chiralities of both multiferroic phases, AF2/4 and high-field AF5, across the domain boundaries.

The spatial transition from the AF2/4 conical spin structure (Fig. \ref{Fig0}b) to the high-field AF5 structure (Fig. \ref{Fig0}c) across their common domain boundary can be visualized by decreasing the axial component of the conical spin vector and by tilting the normal vector $\overrightarrow{n}$ of the spiral towards the $c$-axis, as sketched in Fig. \ref{Fig8}. This transition is smooth and it is most likely that the chirality of the helical spin order is preserved across the domain boundary to minimize the boundary energy. The ferroelectric polarization in both domains is parallel to the $b$-axis, however, contributions from $\overrightarrow{P}^{(J1)}$ and $\overrightarrow{P}^{(J3)}$ are different in the AF5 and AF2/4 domains. According to equations (1) and (2), $P_b$ is the sum of contributions from both exchange pathways, J1 and J3, and can be written as

\begin{eqnarray}
P_b=m_{\parallel}m_{\perp}[C^{(J1)}sin(\pi Q_z)sin\Theta cos\varphi\nonumber \\ -C^{(J3)}sin(2\pi Q_x)cos\Theta]
\end{eqnarray}

$C^{(J1)}$ is significantly larger than $C^{(J3)}$ as can be seen from the different values of $P_a$ and $P_c$ in the AF5 phase (e.g. for x=0.1) in zero magnetic field.\cite{liang:12c,olabarria:12} It is not clear, however, whether or not the two contributions to $P_b$ in equation (3) are of equal sign. If $P_b^{(J1)}$ and $P_b^{(J3)}$ are of opposite sign we can explain the observed polarization reversal for x=0.15 (according to the data of Fig. \ref{Fig1}), as discussed below.

It is essential for the following discussion that domains of different magnetic orders (AF2/4 and AF5) spatially coexist in Mn$_{0.85}$Co$_{0.15}$WO$_4$.\cite{chaudhury:10,ye:12} While cooling in a positive poling electric field, the ferroelectric polarization is first established at $T_{C1}$=10.2 K due to the formation of domains with the AF2/4 conical spin structure. With $\theta$=36$^\circ$ and $\varphi$=0$^\circ$,\cite{ye:12,olabarria:12b} the net $P_b$ includes two contributions according to equation (3). We assume in the following that both contributions have opposite sign. The first term from J1 dominates the second term (J3) because of the larger coefficient $C^{(J1)}$ and it is (positive) aligned with the poling field. The second term, opposite in sign, diminishes the total polarization accordingly to the relatively small value of 3 $\mu C/m^2$ (Fig. \ref{Fig1}). This small value also reflects the reduced volume fraction of the conical AF2/4 phase in the mixture of coexisting AF5 and AF2/4 phases. A schematic presentation of the contributions to $P_b$ is shown in the box of Fig. \ref{Fig8}. The AF2/4 phase contribution to $P_b$ decreases slightly with increasing field $H_c$, consistent with the data for higher doping at x=0.17 (Fig. \ref{Fig2}b).

Upon further decreasing temperature, below $T_{C2}\simeq$7 K, the domains of the AF5 phase develop and, above the critical $H_c$ of the spin-flop transition, add another contribution to $P_b$ which arises from the J3 exchange path only since the spiral axis in this phase is parallel to the $c$-axis. Once the chirality of the magnetic order across the domain boundary is preserved, this contribution to $P_b$ has to be negative, i.e. opposite to the ferroelectric polarization established above $T_{C2}$, as schematically shown in the box of Fig. \ref{Fig8}. This negative contribution dominates the overall polarization and results in the sign change of $P_b$ according to our data shown in Fig. \ref{Fig1}.

From the above discussion an interesting conclusion can be drawn: The coexisting domains with different helical spin orders ($a-c$ spiral AF5 and conical AF2/4) in Mn$_{0.85}$Co$_{0.15}$WO$_4$ are strongly coupled and the smooth transition across their domains walls preserves the chirality of the spin order. While this conclusion is indirect from our macroscopic measurements, it could be verified in experiments that are sensitive to or allow to determine the chirality of the spin spiral or the spacial distribution of the sign of $P_b$. Possible examples of those experiments are polarized neutron scattering\cite{seki:08} as well as second harmonic optical imaging\cite{meier:09} or piezoelectric force microscopy. When the Mn$_{0.85}$Co$_{0.15}$WO$_4$ sample is cooled below $T_{C1}$, the external poling electric field determines the chirality of the multiferroic AF2/4 domains. Upon further cooling, and in an external $c$-axis magnetic field, domains of the second (high-field) multiferroic AF5 phase form at $T_{C2}$. The strong coupling of both helical magnetic orders results in the AF5 order having the same chirality which determines the orientation of the polarization of the AF5 domain in opposite direction to that of the AF2/4 domain. The polarization reversal happens despite the positive poling electric field which is an indication of the strength of the interdomain coupling and the energy gained by preserving the spin chirality across the domain walls. Future work should focus on the study of the properties of the coexisting multiferroic domains and the transition across their respective boundaries.

\section{Conclusions}
In zero magnetic field, the cobalt doped Mn$_{1-x}$Co$_x$WO$_4$ shows one of the most complex phase diagrams with repeated flops of the ferroelectric polarization upon increasing Co content. The application of external magnetic fields adds another dimension to the stability and tunability of multiple multiferroic phases. At a critical concentration x=0.15, cooling in a magnetic field aligned with the $c$-axis results in a spontaneous polarization reversal at $T_{C2}\simeq$ 7 K. The ferroelectric polarization, aligned with the $b$-axis of the crystal ($P_b$), changes sign despite a poling electrical field in the opposite direction. The origin of this sign reversal was found in the specifics of the multiferroic phases of Mn$_{0.85}$Co$_{0.15}$WO$_4$:

(i) Two major magnetic exchange interactions contribute to the ferroelectric polarization, J1 and J3, which label the exchange between nearest neighbors along the $c$-axis and $a$-axis, respectively. Both exchange interactions can contribute to the $b$-axis polarization and their contributions are of opposite sign.

(ii) At zero magnetic field two multiferroic phases coexist in Mn$_{0.85}$Co$_{0.15}$WO$_4$ in form of domains separated by domain walls. The AF2/4 magnetic domain represents the conical spin structure that is known for x$>$0.15. The AF5 domain shows the $a-c$ spin spiral characteristic for Mn$_{1-x}$Co$_x$WO$_4$ with 0.075$<$x$<$0.15. The spin spiral of this phase flops into an $a-b$ spiral above a critical $c$-axis magnetic field, allowing for a contribution to $P_b$ which is superimposed to the polarization arising from the AF2/4 domains.

(iii) The magnetic domains of both multiferroic phases are strongly coupled and the spin chirality is preserved in crossing the domain wall. This strong correlation forces $P_b$ below $T_{C2}$ to be counteraligned with the polarization that originates from the AF2/4 domain below the higher temperature $T_{C1}$, and it explains the sign change of $P_b$ observed experimentally in Mn$_{0.85}$Co$_{0.15}$WO$_4$ above a critical field $H_c$.

The results of this work demonstrate the complexity of this particular multiferroic compound, Mn$_{1-x}$Co$_{x}$WO$_4$, and it shows new ways to control and manipulate the microscopic parameters (exchange coupling, anisotropy), the mesoscopic domain properties, and the macroscopic quantities such as the ferroelectric polarization. Further studies are warranted and suggested to explore and to better understand the domain properties and the correlation across the domains walls on a microscopic level.

\begin{acknowledgments}
This work is supported in part by the US Air Force Office of Scientific Research (AFOSR) Grant No. FA9550-09-1-0656, the T.L.L. Temple Foundation, the John J. and Rebecca Moores Endowment, and the State of Texas through the Texas Center for Superconductivity at the University of Houston. The work at ORNL is partially supported by the DOE BES Office of Scientific User Facilities.
\end{acknowledgments}


\begin{thebibliography}{10}%
\makeatletter
\providecommand \@ifxundefined [1]{%
 \ifx #1\undefined \expandafter \@firstoftwo
 \else \expandafter \@secondoftwo
\fi
}%
\providecommand \@ifnum [1]{%
 \ifnum #1\expandafter \@firstoftwo
 \else \expandafter \@secondoftwo
\fi
}%
\providecommand \enquote [1]{``#1''}%
\providecommand \bibnamefont  [1]{#1}%
\providecommand \bibfnamefont [1]{#1}%
\providecommand \citenamefont [1]{#1}%
\providecommand\href[0]{\@sanitize\@href}%
\providecommand\@href[1]{\endgroup\@@startlink{#1}\endgroup\@@href}%
\providecommand\@@href[1]{#1\@@endlink}%
\providecommand \@sanitize [0]{\begingroup\catcode`\&12\catcode`\#12\relax}%
\@ifxundefined \pdfoutput {\@firstoftwo}{%
 \@ifnum{\z@=\pdfoutput}{\@firstoftwo}{\@secondoftwo}%
}{%
 \providecommand\@@startlink[1]{\leavevmode}%
 \providecommand\@@endlink[0]{}%
}{%
 \providecommand\@@startlink[1]{%
  \leavevmode
  \pdfstartlink
   attr{/Border[0 0 1 ]/H/I/C[0 1 1]}%
   user{/Subtype/Link/A<</Type/Action/S/URI/URI(#1)>>}%
  \relax
 }%
 \providecommand\@@endlink[0]{\pdfendlink}%
}%
\providecommand \url  [0]{\begingroup\@sanitize \@url }%
\providecommand \@url [1]{\endgroup\@href {#1}{\urlprefix}}%
\providecommand \urlprefix [0]{URL }%
\providecommand \Eprint[0]{\href }%
\@ifxundefined \urlstyle {%
  \providecommand \doi [1]{doi:\discretionary{}{}{}#1}%
}{%
  \providecommand \doi [0]{doi:\discretionary{}{}{}\begingroup
  \urlstyle{rm}\Url }%
}%
\providecommand \doibase [0]{http://dx.doi.org/}%
\providecommand \Doi[1]{\href{\doibase#1}}%
\providecommand \bibAnnote [3]{%
  \BibitemShut{#1}%
  \begin{quotation}\noindent
    \textsc{Key:}\ #2\\\textsc{Annotation:}\ #3%
  \end{quotation}%
}%
\providecommand \bibAnnoteFile [2]{%
  \IfFileExists{#2}{\bibAnnote {#1} {#2} {\input{#2}}}{}%
}%
\providecommand \typeout [0]{\immediate \write \m@ne }%
\providecommand \selectlanguage [0]{\@gobble}%
\providecommand \bibinfo [0]{\@secondoftwo}%
\providecommand \bibfield [0]{\@secondoftwo}%
\providecommand \translation [1]{[#1]}%
\providecommand \BibitemOpen[0]{}%
\providecommand \bibitemStop [0]{}%
\providecommand \bibitemNoStop [0]{.\EOS\space}%
\providecommand \EOS [0]{\spacefactor3000\relax}%
\providecommand \BibitemShut [1]{\csname bibitem#1\endcsname}%
\bibitem{kimura:03}%
  \BibitemOpen
  \bibfield{author}{%
  \bibinfo {author} {\bibfnamefont{T.}~\bibnamefont{Kimura}}, \bibinfo {author}
  {\bibfnamefont{T.}~\bibnamefont{Goto}}, \bibinfo {author}
  {\bibfnamefont{H.}~\bibnamefont{Shintani}}, \bibinfo {author}
  {\bibfnamefont{K.}~\bibnamefont{Ishizaka}}, \bibinfo {author}
  {\bibfnamefont{T.}~\bibnamefont{Arima}},\ and\ \bibinfo {author}
  {\bibfnamefont{Y.}~\bibnamefont{Tokura}},\ }%
  \bibfield{journal}{%
  \bibinfo {journal} {Nature (London)}\ }%
  \textbf{\bibinfo {volume} {426}},\ \bibinfo {pages} {55} (\bibinfo {year}
  {2003})%
  \bibAnnoteFile{NoStop}{kimura:03}%
\bibitem{mostovoy:06}%
  \BibitemOpen
  \bibfield{author}{%
  \bibinfo {author} {\bibfnamefont{M.}~\bibnamefont{Mostovoy}},\ }%
  \bibfield{journal}{%
  \bibinfo {journal} {Phys. Rev. Lett.}\ }%
  \textbf{\bibinfo {volume} {96}},\ \bibinfo {pages} {067601} (\bibinfo {year}
  {2006})%
  \bibAnnoteFile{NoStop}{mostovoy:06}%
\bibitem{katsura:05}%
  \BibitemOpen
  \bibfield{author}{%
  \bibinfo {author} {\bibfnamefont{H.}~\bibnamefont{Katsura}}, \bibinfo
  {author} {\bibfnamefont{N.}~\bibnamefont{Nagaosa}},\ and\ \bibinfo {author}
  {\bibfnamefont{A.~V.}\ \bibnamefont{Balatsky}},\ }%
  \bibfield{journal}{%
  \bibinfo {journal} {Phys. Rev. Lett.}\ }%
  \textbf{\bibinfo {volume} {95}},\ \bibinfo {pages} {057205} (\bibinfo {year}
  {2005})%
  \bibAnnoteFile{NoStop}{katsura:05}%
\bibitem{sergienko:06}%
  \BibitemOpen
  \bibfield{author}{%
  \bibinfo {author} {\bibfnamefont{I.~A.}\ \bibnamefont{Sergienko}}\ and\
  \bibinfo {author} {\bibfnamefont{E.}~\bibnamefont{Dagotto}},\ }%
  \bibfield{journal}{%
  \bibinfo {journal} {Phys. Rev. B}\ }%
  \textbf{\bibinfo {volume} {73}},\ \bibinfo {pages} {094434} (\bibinfo {year}
  {2006})%
  \bibAnnoteFile{NoStop}{sergienko:06}%
\bibitem{fiebig:05}%
  \BibitemOpen
  \bibfield{author}{%
  \bibinfo {author} {\bibfnamefont{M.}~\bibnamefont{Fiebig}},\ }%
  \bibfield{journal}{%
  \bibinfo {journal} {J. Phys. D: Appl. Phys.}\ }%
  \textbf{\bibinfo {volume} {38}},\ \bibinfo {pages} {R123} (\bibinfo {year}
  {2005})%
  \bibAnnoteFile{NoStop}{fiebig:05}%
\bibitem{spaldin:05}%
  \BibitemOpen
  \bibfield{author}{%
  \bibinfo {author} {\bibfnamefont{N.~A.}\ \bibnamefont{Spaldin}}\ and\
  \bibinfo {author} {\bibfnamefont{M.}~\bibnamefont{Fiebig}},\ }%
  \bibfield{journal}{%
  \bibinfo {journal} {Science}\ }%
  \textbf{\bibinfo {volume} {309}},\ \bibinfo {pages} {391} (\bibinfo {year}
  {2005})%
  \bibAnnoteFile{NoStop}{spaldin:05}%
\bibitem{tokura:07}%
  \BibitemOpen
  \bibfield{author}{%
  \bibinfo {author} {\bibfnamefont{Y.}~\bibnamefont{Tokura}},\ }%
  \bibfield{journal}{%
  \bibinfo {journal} {J. Mag. Mag. Mat.}\ }%
  \textbf{\bibinfo {volume} {310}},\ \bibinfo {pages} {1145} (\bibinfo {year}
  {2007})%
  \bibAnnoteFile{NoStop}{tokura:07}%
\bibitem{higashiyama:04}%
  \BibitemOpen
  \bibfield{author}{%
  \bibinfo {author} {\bibfnamefont{D.}~\bibnamefont{Higashiyama}}, \bibinfo
  {author} {\bibfnamefont{S.}~\bibnamefont{Miyasaka}}, \bibinfo {author}
  {\bibfnamefont{N.}~\bibnamefont{Kida}}, \bibinfo {author}
  {\bibfnamefont{T.}~\bibnamefont{Arima}},\ and\ \bibinfo {author}
  {\bibfnamefont{Y.}~\bibnamefont{Tokura}},\ }%
  \bibfield{journal}{%
  \bibinfo {journal} {Phys. Rev. B}\ }%
  \textbf{\bibinfo {volume} {70}},\ \bibinfo {pages} {174405} (\bibinfo {year}
  {2004})%
  \bibAnnoteFile{NoStop}{higashiyama:04}%
\bibitem{hur:04}%
  \BibitemOpen
  \bibfield{author}{%
  \bibinfo {author} {\bibfnamefont{N.}~\bibnamefont{Hur}}, \bibinfo {author}
  {\bibfnamefont{S.}~\bibnamefont{Park}}, \bibinfo {author}
  {\bibfnamefont{P.~A.}\ \bibnamefont{Sharma}}, \bibinfo {author}
  {\bibfnamefont{S.}~\bibnamefont{Guha}},\ and\ \bibinfo {author}
  {\bibfnamefont{S.-W.}\ \bibnamefont{Cheong}},\ }%
  \bibfield{journal}{%
  \bibinfo {journal} {Phys. Rev. Lett.}\ }%
  \textbf{\bibinfo {volume} {93}},\ \bibinfo {pages} {107207} (\bibinfo {year}
  {2004})%
  \bibAnnoteFile{NoStop}{hur:04}%
\bibitem{taniguchi:06}%
  \BibitemOpen
  \bibfield{author}{%
  \bibinfo {author} {\bibfnamefont{K.}~\bibnamefont{Taniguchi}}, \bibinfo
  {author} {\bibfnamefont{N.}~\bibnamefont{Abe}}, \bibinfo {author}
  {\bibfnamefont{T.}~\bibnamefont{Takenobu}}, \bibinfo {author}
  {\bibfnamefont{Y.}~\bibnamefont{Iwasa}},\ and\ \bibinfo {author}
  {\bibfnamefont{T.}~\bibnamefont{Arima}},\ }%
  \bibfield{journal}{%
  \bibinfo {journal} {Phys. Rev. Lett.}\ }%
  \textbf{\bibinfo {volume} {97}},\ \bibinfo {pages} {097203} (\bibinfo {year}
  {2006})%
  \bibAnnoteFile{NoStop}{taniguchi:06}%
\bibitem{seki:08}%
  \BibitemOpen
  \bibfield{author}{%
  \bibinfo {author} {\bibfnamefont{S.}~\bibnamefont{Seki}}, \bibinfo {author}
  {\bibfnamefont{Y.}~\bibnamefont{Yamasaki}}, \bibinfo {author}
  {\bibfnamefont{M.}~\bibnamefont{Soda}}, \bibinfo {author}
  {\bibfnamefont{M.}~\bibnamefont{Matsuura}}, \bibinfo {author}
  {\bibfnamefont{K.}~\bibnamefont{Hirota}},\ and\ \bibinfo {author}
  {\bibfnamefont{Y.}~\bibnamefont{Tokura}},\ }%
  \bibfield{journal}{%
  \bibinfo {journal} {Phys. Rev. Lett.}\ }%
  \textbf{\bibinfo {volume} {100}},\ \bibinfo {pages} {127201} (\bibinfo {year}
  {2008})%
  \bibAnnoteFile{NoStop}{seki:08}%
\bibitem{sagayama:08}%
  \BibitemOpen
  \bibfield{author}{%
  \bibinfo {author} {\bibfnamefont{H.}~\bibnamefont{Sagayama}}, \bibinfo
  {author} {\bibfnamefont{K.}~\bibnamefont{Taniguchi}}, \bibinfo {author}
  {\bibfnamefont{N.}~\bibnamefont{Abe}}, \bibinfo {author}
  {\bibfnamefont{T.}~\bibnamefont{Arima}}, \bibinfo {author}
  {\bibfnamefont{M.}~\bibnamefont{Soda}}, \bibinfo {author}
  {\bibfnamefont{M.}~\bibnamefont{Matsuura}},\ and\ \bibinfo {author}
  {\bibfnamefont{K.}~\bibnamefont{Hirota}},\ }%
  \bibfield{journal}{%
  \bibinfo {journal} {Phys. Rev. B}\ }%
  \textbf{\bibinfo {volume} {77}},\ \bibinfo {pages} {220407(R)} (\bibinfo
  {year} {2008})%
  \bibAnnoteFile{NoStop}{sagayama:08}%
\bibitem{delacruz:07}%
  \BibitemOpen
  \bibfield{author}{%
  \bibinfo {author} {\bibfnamefont{C.~R.}\ \bibnamefont{dela Cruz}}, \bibinfo
  {author} {\bibfnamefont{B.}~\bibnamefont{Lorenz}}, \bibinfo {author}
  {\bibfnamefont{Y.~Y.}\ \bibnamefont{Sun}}, \bibinfo {author}
  {\bibfnamefont{Y.}~\bibnamefont{Wang}}, \bibinfo {author}
  {\bibfnamefont{S.}~\bibnamefont{Park}}, \bibinfo {author}
  {\bibfnamefont{S.-W.}\ \bibnamefont{Cheong}}, \bibinfo {author}
  {\bibfnamefont{M.~M.}\ \bibnamefont{Gospodinov}},\ and\ \bibinfo {author}
  {\bibfnamefont{C.~W.}\ \bibnamefont{Chu}},\ }%
  \bibfield{journal}{%
  \bibinfo {journal} {Phys. Rev. B}\ }%
  \textbf{\bibinfo {volume} {76}},\ \bibinfo {pages} {174106} (\bibinfo {year}
  {2007})%
  \bibAnnoteFile{NoStop}{delacruz:07}%
\bibitem{chaudhury:07}%
  \BibitemOpen
  \bibfield{author}{%
  \bibinfo {author} {\bibfnamefont{R.~P.}\ \bibnamefont{Chaudhury}}, \bibinfo
  {author} {\bibfnamefont{F.}~\bibnamefont{Yen}}, \bibinfo {author}
  {\bibfnamefont{C.~R.}\ \bibnamefont{dela Cruz}}, \bibinfo {author}
  {\bibfnamefont{B.}~\bibnamefont{Lorenz}}, \bibinfo {author}
  {\bibfnamefont{Y.~Q.}\ \bibnamefont{Wang}}, \bibinfo {author}
  {\bibfnamefont{Y.~Y.}\ \bibnamefont{Sun}},\ and\ \bibinfo {author}
  {\bibfnamefont{C.~W.}\ \bibnamefont{Chu}},\ }%
  \bibfield{journal}{%
  \bibinfo {journal} {Phys. Rev. B}\ }%
  \textbf{\bibinfo {volume} {75}},\ \bibinfo {pages} {012407} (\bibinfo {year}
  {2007})%
  \bibAnnoteFile{NoStop}{chaudhury:07}%
\bibitem{chaudhury:08}%
  \BibitemOpen
  \bibfield{author}{%
  \bibinfo {author} {\bibfnamefont{R.~P.}\ \bibnamefont{Chaudhury}}, \bibinfo
  {author} {\bibfnamefont{B.}~\bibnamefont{Lorenz}}, \bibinfo {author}
  {\bibfnamefont{Y.~Q.}\ \bibnamefont{Wang}}, \bibinfo {author}
  {\bibfnamefont{Y.~Y.}\ \bibnamefont{Sun}},\ and\ \bibinfo {author}
  {\bibfnamefont{C.~W.}\ \bibnamefont{Chu}},\ }%
  \bibfield{journal}{%
  \bibinfo {journal} {Phys. Rev. B}\ }%
  \textbf{\bibinfo {volume} {77}},\ \bibinfo {pages} {104406} (\bibinfo {year}
  {2008})%
  \bibAnnoteFile{NoStop}{chaudhury:08}%
\bibitem{delacruz:08b}%
  \BibitemOpen
  \bibfield{author}{%
  \bibinfo {author} {\bibfnamefont{C.~R.}\ \bibnamefont{dela Cruz}}, \bibinfo
  {author} {\bibfnamefont{B.}~\bibnamefont{Lorenz}}, \bibinfo {author}
  {\bibfnamefont{W.}~\bibnamefont{Ratcliff}}, \bibinfo {author}
  {\bibfnamefont{J.}~\bibnamefont{Lynn}}, \bibinfo {author}
  {\bibfnamefont{M.~M.}\ \bibnamefont{Gospodinov}},\ and\ \bibinfo {author}
  {\bibfnamefont{C.~W.}\ \bibnamefont{Chu}},\ }%
  \bibfield{journal}{%
  \bibinfo {journal} {Physica B}\ }%
  \textbf{\bibinfo {volume} {403}},\ \bibinfo {pages} {1359} (\bibinfo {year}
  {2008})%
  \bibAnnoteFile{NoStop}{delacruz:08b}%
\bibitem{seki:07}%
  \BibitemOpen
  \bibfield{author}{%
  \bibinfo {author} {\bibfnamefont{S.}~\bibnamefont{Seki}}, \bibinfo {author}
  {\bibfnamefont{Y.}~\bibnamefont{Yamasaki}}, \bibinfo {author}
  {\bibfnamefont{Y.}~\bibnamefont{Shiomi}}, \bibinfo {author}
  {\bibfnamefont{S.}~\bibnamefont{Iguchi}}, \bibinfo {author}
  {\bibfnamefont{Y.}~\bibnamefont{Onose}},\ and\ \bibinfo {author}
  {\bibfnamefont{Y.}~\bibnamefont{Tokura}},\ }%
  \bibfield{journal}{%
  \bibinfo {journal} {Phys. Rev. B}\ }%
  \textbf{\bibinfo {volume} {75}},\ \bibinfo {pages} {100403(R)} (\bibinfo
  {year} {2007})%
  \bibAnnoteFile{NoStop}{seki:07}%
\bibitem{kanetsuki:07}%
  \BibitemOpen
  \bibfield{author}{%
  \bibinfo {author} {\bibfnamefont{S.}~\bibnamefont{Kanetsuki}}, \bibinfo
  {author} {\bibfnamefont{S.}~\bibnamefont{Mitsuda}}, \bibinfo {author}
  {\bibfnamefont{T.}~\bibnamefont{Nakajima}}, \bibinfo {author}
  {\bibfnamefont{D.}~\bibnamefont{Anazawa}}, \bibinfo {author}
  {\bibfnamefont{H.~A.}\ \bibnamefont{Katori}},\ and\ \bibinfo {author}
  {\bibfnamefont{K.}~\bibnamefont{Prokes}},\ }%
  \bibfield{journal}{%
  \bibinfo {journal} {J. Phys.: Condens. Matter}\ }%
  \textbf{\bibinfo {volume} {19}},\ \bibinfo {pages} {145244} (\bibinfo {year}
  {2007})%
  \bibAnnoteFile{NoStop}{kanetsuki:07}%
\bibitem{kenzelmann:05}%
  \BibitemOpen
  \bibfield{author}{%
  \bibinfo {author} {\bibfnamefont{M.}~\bibnamefont{Kenzelmann}}, \bibinfo
  {author} {\bibfnamefont{A.~B.}\ \bibnamefont{Harris}}, \bibinfo {author}
  {\bibfnamefont{S.}~\bibnamefont{Jonas}}, \bibinfo {author}
  {\bibfnamefont{C.}~\bibnamefont{Broholm}}, \bibinfo {author}
  {\bibfnamefont{J.}~\bibnamefont{Schefer}}, \bibinfo {author}
  {\bibfnamefont{S.~B.}\ \bibnamefont{Kim}}, \bibinfo {author}
  {\bibfnamefont{C.~L.}\ \bibnamefont{Zhang}}, \bibinfo {author}
  {\bibfnamefont{S.-W.}\ \bibnamefont{Cheong}}, \bibinfo {author}
  {\bibfnamefont{O.~P.}\ \bibnamefont{Vajk}},\ and\ \bibinfo {author}
  {\bibfnamefont{J.~W.}\ \bibnamefont{Lynn}},\ }%
  \bibfield{journal}{%
  \bibinfo {journal} {Phys. Rev. Lett.}\ }%
  \textbf{\bibinfo {volume} {95}},\ \bibinfo {pages} {087206} (\bibinfo {year}
  {2005})%
  \bibAnnoteFile{NoStop}{kenzelmann:05}%
\bibitem{lawes:05}%
  \BibitemOpen
  \bibfield{author}{%
  \bibinfo {author} {\bibfnamefont{G.}~\bibnamefont{Lawes}}, \bibinfo {author}
  {\bibfnamefont{A.~B.}\ \bibnamefont{Harris}}, \bibinfo {author}
  {\bibfnamefont{T.}~\bibnamefont{Kimura}}, \bibinfo {author}
  {\bibfnamefont{N.}~\bibnamefont{Rogado}}, \bibinfo {author}
  {\bibfnamefont{R.~J.}\ \bibnamefont{Cava}}, \bibinfo {author}
  {\bibfnamefont{A.}~\bibnamefont{Aharony}}, \bibinfo {author}
  {\bibfnamefont{O.}~\bibnamefont{Entin-Wohlman}}, \bibinfo {author}
  {\bibfnamefont{T.}~\bibnamefont{Yildirim}}, \bibinfo {author}
  {\bibfnamefont{M.}~\bibnamefont{Kenzelmann}}, \bibinfo {author}
  {\bibfnamefont{C.}~\bibnamefont{Broholm}},\ and\ \bibinfo {author}
  {\bibfnamefont{A.~P.}\ \bibnamefont{Ramirez}},\ }%
  \bibfield{journal}{%
  \bibinfo {journal} {Phys. Rev. Lett.}\ }%
  \textbf{\bibinfo {volume} {95}},\ \bibinfo {pages} {087205} (\bibinfo {year}
  {2005})%
  \bibAnnoteFile{NoStop}{lawes:05}%
\bibitem{yasui:08}%
  \BibitemOpen
  \bibfield{author}{%
  \bibinfo {author} {\bibfnamefont{Y.}~\bibnamefont{Yasui}}, \bibinfo {author}
  {\bibfnamefont{Y.}~\bibnamefont{Naito}}, \bibinfo {author}
  {\bibfnamefont{K.}~\bibnamefont{Sato}}, \bibinfo {author}
  {\bibfnamefont{T.}~\bibnamefont{Moyoshi}}, \bibinfo {author}
  {\bibfnamefont{M.}~\bibnamefont{Sato}},\ and\ \bibinfo {author}
  {\bibfnamefont{K.}~\bibnamefont{Kakurai}},\ }%
  \bibfield{journal}{%
  \bibinfo {journal} {J. Phys. Soc. Jpn.}\ }%
  \textbf{\bibinfo {volume} {77}},\ \bibinfo {pages} {023712} (\bibinfo {year}
  {2008})%
  \bibAnnoteFile{NoStop}{yasui:08}%
\bibitem{yamasaki:06}%
  \BibitemOpen
  \bibfield{author}{%
  \bibinfo {author} {\bibfnamefont{Y.}~\bibnamefont{Yamasaki}}, \bibinfo
  {author} {\bibfnamefont{S.}~\bibnamefont{Miyasaki}}, \bibinfo {author}
  {\bibfnamefont{Y.}~\bibnamefont{Kaneko}}, \bibinfo {author}
  {\bibfnamefont{J.-P.}\ \bibnamefont{He}}, \bibinfo {author}
  {\bibfnamefont{T.}~\bibnamefont{Arima}},\ and\ \bibinfo {author}
  {\bibfnamefont{Y.}~\bibnamefont{Tokura}},\ }%
  \bibfield{journal}{%
  \bibinfo {journal} {Phys. Rev. Lett.}\ }%
  \textbf{\bibinfo {volume} {96}},\ \bibinfo {pages} {207204} (\bibinfo {year}
  {2006})%
  \bibAnnoteFile{NoStop}{yamasaki:06}%
\bibitem{arkenbout:06}%
  \BibitemOpen
  \bibfield{author}{%
  \bibinfo {author} {\bibfnamefont{A.~H.}\ \bibnamefont{Arkenbout}}, \bibinfo
  {author} {\bibfnamefont{T.~T.~M.}\ \bibnamefont{Palstra}}, \bibinfo {author}
  {\bibfnamefont{T.}~\bibnamefont{Siegrist}},\ and\ \bibinfo {author}
  {\bibfnamefont{T.}~\bibnamefont{Kimura}},\ }%
  \bibfield{journal}{%
  \bibinfo {journal} {Phys. Rev. B}\ }%
  \textbf{\bibinfo {volume} {74}},\ \bibinfo {pages} {184431} (\bibinfo {year}
  {2006})%
  \bibAnnoteFile{NoStop}{arkenbout:06}%
\bibitem{agricola:1546}%
  \BibitemOpen
  \bibfield{author}{%
  \bibinfo {author} {\bibfnamefont{G.}~\bibnamefont{Agricola}},\ }%
  \emph{\bibinfo {title} {De Natura Fossillium}}, 1546\ (\bibinfo {publisher}
  {engl. transl: M. C. Bandy and J. Bandy, The Geological Society of America},\ \bibinfo {address} {New York}, 1955\bibinfo
  {year})%
  \bibAnnoteFile{NoStop}{agricola:1546}%
\bibitem{lautenschlager:93}%
  \BibitemOpen
  \bibfield{author}{%
  \bibinfo {author} {\bibfnamefont{G.}~\bibnamefont{Lautenschl{\"a}ger}},
  \bibinfo {author} {\bibfnamefont{H.}~\bibnamefont{Weitzel}}, \bibinfo
  {author} {\bibfnamefont{T.}~\bibnamefont{Vogt}}, \bibinfo {author}
  {\bibfnamefont{R.}~\bibnamefont{Hock}}, \bibinfo {author}
  {\bibfnamefont{A.}~\bibnamefont{Bohm}}, \bibinfo {author}
  {\bibfnamefont{M.}~\bibnamefont{Bonnet}},\ and\ \bibinfo {author}
  {\bibfnamefont{H.}~\bibnamefont{Fuess}},\ }%
  \bibfield{journal}{%
  \bibinfo {journal} {Phys. Rev. B}\ }%
  \textbf{\bibinfo {volume} {48}},\ \bibinfo {pages} {6087} (\bibinfo {year}
  {1993})%
  \bibAnnoteFile{NoStop}{lautenschlager:93}%
\bibitem{weitzel:70}%
  \BibitemOpen
  \bibfield{author}{%
  \bibinfo {author} {\bibfnamefont{H.}~\bibnamefont{Weitzel}},\ }%
  \bibfield{journal}{%
  \bibinfo {journal} {Solid State Commun.}\ }%
  \textbf{\bibinfo {volume} {8}},\ \bibinfo {pages} {2071} (\bibinfo {year}
  {1970})%
  \bibAnnoteFile{NoStop}{weitzel:70}%
\bibitem{kleykamp:80}%
  \BibitemOpen
  \bibfield{author}{%
  \bibinfo {author} {\bibfnamefont{H.}~\bibnamefont{Kleykamp}},\ }%
  \bibfield{journal}{%
  \bibinfo {journal} {J. Less-Common Metals}\ }%
  \textbf{\bibinfo {volume} {71}},\ \bibinfo {pages} {127} (\bibinfo {year}
  {1980})%
  \bibAnnoteFile{NoStop}{kleykamp:80}%
\bibitem{takagi:81}%
  \BibitemOpen
  \bibfield{author}{%
  \bibinfo {author} {\bibfnamefont{K.}~\bibnamefont{Takagi}}, \bibinfo {author}
  {\bibfnamefont{T.}~\bibnamefont{Oi}},\ and\ \bibinfo {author}
  {\bibfnamefont{T.}~\bibnamefont{Fukazawa}},\ }%
  \bibfield{journal}{%
  \bibinfo {journal} {J. Crystal Growth}\ }%
  \textbf{\bibinfo {volume} {52}},\ \bibinfo {pages} {580} (\bibinfo {year}
  {1981})%
  \bibAnnoteFile{NoStop}{takagi:81}%
\bibitem{arora:88}%
  \BibitemOpen
  \bibfield{author}{%
  \bibinfo {author} {\bibfnamefont{S.~K.}\ \bibnamefont{Arora}}, \bibinfo
  {author} {\bibfnamefont{T.}~\bibnamefont{Mathew}},\ and\ \bibinfo {author}
  {\bibfnamefont{N.~M.}\ \bibnamefont{Batra}},\ }%
  \bibfield{journal}{%
  \bibinfo {journal} {J. Crystal Growth}\ }%
  \textbf{\bibinfo {volume} {88}},\ \bibinfo {pages} {379} (\bibinfo {year}
  {1988})%
  \bibAnnoteFile{NoStop}{arora:88}%
\bibitem{meddar:09}%
  \BibitemOpen
  \bibfield{author}{%
  \bibinfo {author} {\bibfnamefont{L.}~\bibnamefont{Meddar}}, \bibinfo {author}
  {\bibfnamefont{M.}~\bibnamefont{Josse}}, \bibinfo {author}
  {\bibfnamefont{P.}~\bibnamefont{Deniard}}, \bibinfo {author}
  {\bibfnamefont{C.}~\bibnamefont{La}}, \bibinfo {author}
  {\bibfnamefont{G.}~\bibnamefont{Andre}}, \bibinfo {author}
  {\bibfnamefont{F.}~\bibnamefont{Damay}}, \bibinfo {author}
  {\bibfnamefont{V.}~\bibnamefont{Petricek}}, \bibinfo {author}
  {\bibfnamefont{S.}~\bibnamefont{Jobic}}, \bibinfo {author}
  {\bibfnamefont{M.-H.}\ \bibnamefont{Whangbo}}, \bibinfo {author}
  {\bibfnamefont{M.}~\bibnamefont{Maglione}},\ and\ \bibinfo {author}
  {\bibfnamefont{C.}~\bibnamefont{Payen}},\ }%
  \bibfield{journal}{%
  \bibinfo {journal} {Chem. Mat.}\ }%
  \textbf{\bibinfo {volume} {21}},\ \bibinfo {pages} {5203} (\bibinfo {year}
  {2009})%
  \bibAnnoteFile{NoStop}{meddar:09}%
\bibitem{chaudhury:11}%
  \BibitemOpen
  \bibfield{author}{%
  \bibinfo {author} {\bibfnamefont{R.~P.}\ \bibnamefont{Chaudhury}}, \bibinfo
  {author} {\bibfnamefont{F.}~\bibnamefont{Ye}}, \bibinfo {author}
  {\bibfnamefont{J.~A.}\ \bibnamefont{Fernandez-Baca}}, \bibinfo {author}
  {\bibfnamefont{B.}~\bibnamefont{Lorenz}}, \bibinfo {author}
  {\bibfnamefont{Y.~Q.}\ \bibnamefont{Wang}}, \bibinfo {author}
  {\bibfnamefont{Y.~Y.}\ \bibnamefont{Sun}}, \bibinfo {author}
  {\bibfnamefont{H.~A.}\ \bibnamefont{Mook}},\ and\ \bibinfo {author}
  {\bibfnamefont{C.~W.}\ \bibnamefont{Chu}},\ }%
  \bibfield{journal}{%
  \bibinfo {journal} {Phys. Rev. B}\ }%
  \textbf{\bibinfo {volume} {83}},\ \bibinfo {pages} {014401} (\bibinfo {year}
  {2011})%
  \bibAnnoteFile{NoStop}{chaudhury:11}%
\bibitem{garciamatres:03}%
  \BibitemOpen
  \bibfield{author}{%
  \bibinfo {author} {\bibfnamefont{E.}~\bibnamefont{Garc\'{\i}a-Matres}},
  \bibinfo {author} {\bibfnamefont{N.}~\bibnamefont{St{\"u}${\ss}$er}},
  \bibinfo {author} {\bibfnamefont{M.}~\bibnamefont{Hofmann}},\ and\ \bibinfo
  {author} {\bibfnamefont{M.}~\bibnamefont{Reehuis}},\ }%
  \bibfield{journal}{%
  \bibinfo {journal} {Eur. Phys. J. B}\ }%
  \textbf{\bibinfo {volume} {32}},\ \bibinfo {pages} {35} (\bibinfo {year}
  {2003})%
  \bibAnnoteFile{NoStop}{garciamatres:03}%
\bibitem{chaudhury:09b}%
  \BibitemOpen
  \bibfield{author}{%
  \bibinfo {author} {\bibfnamefont{R.~P.}\ \bibnamefont{Chaudhury}}, \bibinfo
  {author} {\bibfnamefont{B.}~\bibnamefont{Lorenz}}, \bibinfo {author}
  {\bibfnamefont{Y.-Q.}\ \bibnamefont{Wang}}, \bibinfo {author}
  {\bibfnamefont{Y.~Y.}\ \bibnamefont{Sun}},\ and\ \bibinfo {author}
  {\bibfnamefont{C.~W.}\ \bibnamefont{Chu}},\ }%
  \bibfield{journal}{%
  \bibinfo {journal} {New J. Phys.}\ }%
  \textbf{\bibinfo {volume} {11}},\ \bibinfo {pages} {033036} (\bibinfo {year}
  {2009})%
  \bibAnnoteFile{NoStop}{chaudhury:09b}%
\bibitem{matityahu:12}%
  \BibitemOpen
  \bibfield{author}{%
  \bibinfo {author} {\bibfnamefont{S.}~\bibnamefont{Matityahu}}, \bibinfo
  {author} {\bibfnamefont{A.}~\bibnamefont{Aharony}},\ and\ \bibinfo {author}
  {\bibfnamefont{O.}~\bibnamefont{Entin-Wohlman}},\ }%
  \bibfield{journal}{%
  \bibinfo {journal} {Phys. Rev. B}\ }%
  \textbf{\bibinfo {volume} {85}},\ \bibinfo {pages} {174408} (\bibinfo {year}
  {2012})%
  \bibAnnoteFile{NoStop}{matityahu:12}%
\bibitem{liang:12c}%
  \BibitemOpen
  \bibfield{author}{%
  \bibinfo {author} {\bibfnamefont{K.-C.}\ \bibnamefont{Liang}}, \bibinfo
  {author} {\bibfnamefont{Y.-Q.}\ \bibnamefont{Wang}}, \bibinfo {author}
  {\bibfnamefont{Y.~Y.}\ \bibnamefont{Sun}}, \bibinfo {author}
  {\bibfnamefont{B.}~\bibnamefont{Lorenz}}, \bibinfo {author}
  {\bibfnamefont{F.}~\bibnamefont{Ye}}, \bibinfo {author}
  {\bibfnamefont{J.~A.}\ \bibnamefont{Fernandez-Baca}}, \bibinfo {author}
  {\bibfnamefont{H.~A.}\ \bibnamefont{Mook}},\ and\ \bibinfo {author}
  {\bibfnamefont{C.~W.}\ \bibnamefont{Chu}},\ }%
  \bibfield{journal}{%
  \bibinfo {journal} {New J. Phys.}\ }%
  \textbf{\bibinfo {volume} {14}},\ \bibinfo {pages} {073028} (\bibinfo {year}
  {2012})%
  \bibAnnoteFile{NoStop}{liang:12c}%
\bibitem{ye:12}%
  \BibitemOpen
  \bibfield{author}{%
  \bibinfo {author} {\bibfnamefont{F.}~\bibnamefont{Ye}}, \bibinfo {author}
  {\bibfnamefont{S.}~\bibnamefont{Chi}}, \bibinfo {author}
  {\bibfnamefont{J.~A.}\ \bibnamefont{Fernandez-Baca}}, \bibinfo {author}
  {\bibfnamefont{H.}~\bibnamefont{Cao}}, \bibinfo {author}
  {\bibfnamefont{K.-C.}\ \bibnamefont{Liang}}, \bibinfo {author}
  {\bibfnamefont{Y.~Q.}\ \bibnamefont{Wang}}, \bibinfo {author}
  {\bibfnamefont{B.}~\bibnamefont{Lorenz}},\ and\ \bibinfo {author}
  {\bibfnamefont{C.~W.}\ \bibnamefont{Chu}},\ }%
  \bibfield{journal}{%
  \bibinfo {journal} {Phys. Rev. B}\ }%
  \textbf{\bibinfo {volume} {86}},\ \bibinfo {pages} {094429} (\bibinfo {year}
  {2012})%
  \bibAnnoteFile{NoStop}{ye:12}%
\bibitem{weitzel:77}%
  \BibitemOpen
  \bibfield{author}{%
  \bibinfo {author} {\bibfnamefont{H.}~\bibnamefont{Weitzel}}\ and\ \bibinfo
  {author} {\bibfnamefont{H.}~\bibnamefont{Langhof}},\ }%
  \bibfield{journal}{%
  \bibinfo {journal} {J. Mag. Mag. Mat.}\ }%
  \textbf{\bibinfo {volume} {4}},\ \bibinfo {pages} {265} (\bibinfo {year}
  {1977})%
  \bibAnnoteFile{NoStop}{weitzel:77}%
\bibitem{forsyth:94}%
  \BibitemOpen
  \bibfield{author}{%
  \bibinfo {author} {\bibfnamefont{J.~B.}\ \bibnamefont{Forsyth}}\ and\
  \bibinfo {author} {\bibfnamefont{C.}~\bibnamefont{Wilkinson}},\ }%
  \bibfield{journal}{%
  \bibinfo {journal} {J. Phys.: Condens. Matter}\ }%
  \textbf{\bibinfo {volume} {6}},\ \bibinfo {pages} {3073} (\bibinfo {year}
  {1994})%
  \bibAnnoteFile{NoStop}{forsyth:94}%
\bibitem{song:09}%
  \BibitemOpen
  \bibfield{author}{%
  \bibinfo {author} {\bibfnamefont{Y.-S.}\ \bibnamefont{Song}}, \bibinfo
  {author} {\bibfnamefont{J.-H.}\ \bibnamefont{Chung}}, \bibinfo {author}
  {\bibfnamefont{J.~M.~S.}\ \bibnamefont{Park}},\ and\ \bibinfo {author}
  {\bibfnamefont{Y.-N.}\ \bibnamefont{Choi}},\ }%
  \bibfield{journal}{%
  \bibinfo {journal} {Phys. Rev. B}\ }%
  \textbf{\bibinfo {volume} {79}},\ \bibinfo {pages} {224415} (\bibinfo {year}
  {2009})%
  \bibAnnoteFile{NoStop}{song:09}%
\bibitem{song:10}%
  \BibitemOpen
  \bibfield{author}{%
  \bibinfo {author} {\bibfnamefont{Y.-S.}\ \bibnamefont{Song}}, \bibinfo
  {author} {\bibfnamefont{L.~Q.}\ \bibnamefont{Yan}}, \bibinfo {author}
  {\bibfnamefont{B.}~\bibnamefont{Lee}}, \bibinfo {author}
  {\bibfnamefont{S.~H.}\ \bibnamefont{Chun}}, \bibinfo {author}
  {\bibfnamefont{K.~H.}\ \bibnamefont{Kim}}, \bibinfo {author}
  {\bibfnamefont{S.~B.}\ \bibnamefont{Kim}}, \bibinfo {author}
  {\bibfnamefont{A.}~\bibnamefont{Nogami}}, \bibinfo {author}
  {\bibfnamefont{T.}~\bibnamefont{Katsufuji}}, \bibinfo {author}
  {\bibfnamefont{J.}~\bibnamefont{Schefer}},\ and\ \bibinfo {author}
  {\bibfnamefont{J.-H.}\ \bibnamefont{Chung}},\ }%
  \bibfield{journal}{%
  \bibinfo {journal} {Phys. Rev. B}\ }%
  \textbf{\bibinfo {volume} {82}},\ \bibinfo {pages} {214418} (\bibinfo {year}
  {2010})%
  \bibAnnoteFile{NoStop}{song:10}%
\bibitem{olabarria:12}%
  \BibitemOpen
  \bibfield{author}{%
  \bibinfo {author} {\bibfnamefont{I.}~\bibnamefont{Urcelay-Olabarria}},
  \bibinfo {author} {\bibfnamefont{E.}~\bibnamefont{Ressouche}}, \bibinfo
  {author} {\bibfnamefont{A.~A.}\ \bibnamefont{Mukhin}}, \bibinfo {author}
  {\bibfnamefont{Y.~Y.}\ \bibnamefont{Ivanov}}, \bibinfo {author}
  {\bibfnamefont{A.~M.}\ \bibnamefont{Balbashov}}, \bibinfo {author}
  {\bibfnamefont{G.~P.}\ \bibnamefont{Vorobev}}, \bibinfo {author}
  {\bibfnamefont{Y.~F.}\ \bibnamefont{Popov}}, \bibinfo {author}
  {\bibfnamefont{A.~M.}\ \bibnamefont{Kadomtseva}}, \bibinfo {author}
  {\bibfnamefont{J.~L.}\ \bibnamefont{Garcia-Munoz}},\ and\ \bibinfo {author}
  {\bibfnamefont{V.}~\bibnamefont{Skumryev}},\ }%
  \bibfield{journal}{%
  \bibinfo {journal} {Phys. Rev. B}\ }%
  \textbf{\bibinfo {volume} {85}},\ \bibinfo {pages} {094436} (\bibinfo {year}
  {2012})%
  \bibAnnoteFile{NoStop}{olabarria:12}%
\bibitem{olabarria:12b}%
  \BibitemOpen
  \bibfield{author}{%
  \bibinfo {author} {\bibfnamefont{I.}~\bibnamefont{Urcelay-Olabarria}},
  \bibinfo {author} {\bibfnamefont{E.}~\bibnamefont{Ressouche}}, \bibinfo
  {author} {\bibfnamefont{A.~A.}\ \bibnamefont{Mukhin}}, \bibinfo {author}
  {\bibfnamefont{V.~Y.}\ \bibnamefont{Ivanov}}, \bibinfo {author}
  {\bibfnamefont{A.~M.}\ \bibnamefont{Balbashov}}, \bibinfo {author}
  {\bibfnamefont{J.~L.}\ \bibnamefont{Garc\'{\i}a-Mu{\~n}oz}},\ and\ \bibinfo
  {author} {\bibfnamefont{V.}~\bibnamefont{Skumryev}},\ }%
  \bibfield{journal}{%
  \bibinfo {journal} {Phys. Rev. B}\ }%
  \textbf{\bibinfo {volume} {85}},\ \bibinfo {pages} {224419} (\bibinfo {year}
  {2012})%
  \bibAnnoteFile{NoStop}{olabarria:12b}%
\bibitem{chaudhury:10}%
  \BibitemOpen
  \bibfield{author}{%
  \bibinfo {author} {\bibfnamefont{R.~P.}\ \bibnamefont{Chaudhury}}, \bibinfo
  {author} {\bibfnamefont{F.}~\bibnamefont{Ye}}, \bibinfo {author}
  {\bibfnamefont{J.~A.}\ \bibnamefont{Fernandez-Baca}}, \bibinfo {author}
  {\bibfnamefont{Y.-Q.}\ \bibnamefont{Wang}}, \bibinfo {author}
  {\bibfnamefont{Y.~Y.}\ \bibnamefont{Sun}}, \bibinfo {author}
  {\bibfnamefont{B.}~\bibnamefont{Lorenz}}, \bibinfo {author}
  {\bibfnamefont{H.~A.}\ \bibnamefont{Mook}},\ and\ \bibinfo {author}
  {\bibfnamefont{C.~W.}\ \bibnamefont{Chu}},\ }%
  \bibfield{journal}{%
  \bibinfo {journal} {Phys. Rev. B}\ }%
  \textbf{\bibinfo {volume} {82}},\ \bibinfo {pages} {184422} (\bibinfo {year}
  {2010})%
  \bibAnnoteFile{NoStop}{chaudhury:10}%
\bibitem{liang:12}%
  \BibitemOpen
  \bibfield{author}{%
  \bibinfo {author} {\bibfnamefont{K.-C.}\ \bibnamefont{Liang}}, \bibinfo
  {author} {\bibfnamefont{R.~P.}\ \bibnamefont{Chaudhury}}, \bibinfo {author}
  {\bibfnamefont{Y.~Q.}\ \bibnamefont{Wang}}, \bibinfo {author}
  {\bibfnamefont{Y.~Y.}\ \bibnamefont{Sun}}, \bibinfo {author}
  {\bibfnamefont{B.}~\bibnamefont{Lorenz}},\ and\ \bibinfo {author}
  {\bibfnamefont{C.~W.}\ \bibnamefont{Chu}},\ }%
  \bibfield{journal}{%
  \bibinfo {journal} {J. Appl. Phys.}\ }%
  \textbf{\bibinfo {volume} {111}},\ \bibinfo {pages} {07D903} (\bibinfo {year}
  {2012})%
  \bibAnnoteFile{NoStop}{liang:12}%
\bibitem{mitamura:12}%
  \BibitemOpen
  \bibfield{author}{%
  \bibinfo {author} {\bibfnamefont{H.}~\bibnamefont{Mitamura}}, \bibinfo
  {author} {\bibfnamefont{T.}~\bibnamefont{Sakakibara}}, \bibinfo {author}
  {\bibfnamefont{H.}~\bibnamefont{Nakamura}}, \bibinfo {author}
  {\bibfnamefont{T.}~\bibnamefont{Kimura}},\ and\ \bibinfo {author}
  {\bibfnamefont{K.}~\bibnamefont{Kindo}},\ }%
  \bibfield{journal}{%
  \bibinfo {journal} {J. Phys. Soc. Jpn.}\ }%
  \textbf{\bibinfo {volume} {81}},\ \bibinfo {pages} {054705} (\bibinfo {year}
  {2012})%
  \bibAnnoteFile{NoStop}{mitamura:12}%
\bibitem{quirion:13}%
  \BibitemOpen
  \bibfield{author}{%
  \bibinfo {author} {\bibfnamefont{G.}~\bibnamefont{Quirion}}\ and\ \bibinfo
  {author} {\bibfnamefont{M.~L.}\ \bibnamefont{Plumer}},\ }%
  \bibfield{journal}{%
  \bibinfo {journal} {Phys. Rev. B}\ }%
  \textbf{\bibinfo {volume} {87}},\ \bibinfo {pages} {174428} (\bibinfo {year}
  {2013})%
  \bibAnnoteFile{NoStop}{quirion:13}%
\bibitem{stinchcombe:73}%
  \BibitemOpen
  \bibfield{author}{%
  \bibinfo {author} {\bibfnamefont{R.~B.}\ \bibnamefont{Stinchcombe}},\ }%
  \bibfield{journal}{%
  \bibinfo {journal} {J. Physics C: Solid State Phys.}\ }%
  \textbf{\bibinfo {volume} {6}},\ \bibinfo {pages} {2459} (\bibinfo {year}
  {1973})%
  \bibAnnoteFile{NoStop}{stinchcombe:73}%
\bibitem{nagamiya:62}%
  \BibitemOpen
  \bibfield{author}{%
  \bibinfo {author} {\bibfnamefont{T.}~\bibnamefont{Nagamiya}}, \bibinfo
  {author} {\bibfnamefont{K.}~\bibnamefont{Nagata}},\ and\ \bibinfo {author}
  {\bibfnamefont{Y.}~\bibnamefont{Kitano}},\ }%
  \bibfield{journal}{%
  \bibinfo {journal} {Prog. Theor. Phys.}\ }%
  \textbf{\bibinfo {volume} {27}},\ \bibinfo {pages} {1253} (\bibinfo {year}
  {1962})%
  \bibAnnoteFile{NoStop}{nagamiya:62}%
\bibitem{kitano:64}%
  \BibitemOpen
  \bibfield{author}{%
  \bibinfo {author} {\bibfnamefont{Y.}~\bibnamefont{Kitano}}\ and\ \bibinfo
  {author} {\bibfnamefont{T.}~\bibnamefont{Nagamiya}},\ }%
  \bibfield{journal}{%
  \bibinfo {journal} {Prog. Theor. Phys.}\ }%
  \textbf{\bibinfo {volume} {31}},\ \bibinfo {pages} {1} (\bibinfo {year}
  {1964})%
  \bibAnnoteFile{NoStop}{kitano:64}%
\bibitem{tian:09}%
  \BibitemOpen
  \bibfield{author}{%
  \bibinfo {author} {\bibfnamefont{C.}~\bibnamefont{Tian}}, \bibinfo {author}
  {\bibfnamefont{C.}~\bibnamefont{Lee}}, \bibinfo {author}
  {\bibfnamefont{H.}~\bibnamefont{Xiang}}, \bibinfo {author}
  {\bibfnamefont{Y.}~\bibnamefont{Zhang}}, \bibinfo {author}
  {\bibfnamefont{C.}~\bibnamefont{Payen}}, \bibinfo {author}
  {\bibfnamefont{S.}~\bibnamefont{Jobic}},\ and\ \bibinfo {author}
  {\bibfnamefont{M.-H.}\ \bibnamefont{Whangbo}},\ }%
  \bibfield{journal}{%
  \bibinfo {journal} {Phys. Rev. B}\ }%
  \textbf{\bibinfo {volume} {80}},\ \bibinfo {pages} {104426} (\bibinfo {year}
  {2009})%
  \bibAnnoteFile{NoStop}{tian:09}%
\bibitem{ehrenberg:99}%
  \BibitemOpen
  \bibfield{author}{%
  \bibinfo {author} {\bibfnamefont{H.}~\bibnamefont{Ehrenberg}}, \bibinfo
  {author} {\bibfnamefont{H.}~\bibnamefont{Weitzel}}, \bibinfo {author}
  {\bibfnamefont{H.}~\bibnamefont{Fuess}},\ and\ \bibinfo {author}
  {\bibfnamefont{B.}~\bibnamefont{Hennion}},\ }%
  \bibfield{journal}{%
  \bibinfo {journal} {J. Phys. Condens. Matter}\ }%
  \textbf{\bibinfo {volume} {11}},\ \bibinfo {pages} {2649} (\bibinfo {year}
  {1999})%
  \bibAnnoteFile{NoStop}{ehrenberg:99}%
\bibitem{ye:11}%
  \BibitemOpen
  \bibfield{author}{%
  \bibinfo {author} {\bibfnamefont{F.}~\bibnamefont{Ye}}, \bibinfo {author}
  {\bibfnamefont{R.~S.}\ \bibnamefont{Fishman}}, \bibinfo {author}
  {\bibfnamefont{J.~A.}\ \bibnamefont{Fernandez-Baca}}, \bibinfo {author}
  {\bibfnamefont{A.~A.}\ \bibnamefont{Podlesnyak}}, \bibinfo {author}
  {\bibfnamefont{G.}~\bibnamefont{Ehlers}}, \bibinfo {author}
  {\bibfnamefont{H.~A.}\ \bibnamefont{Mook}}, \bibinfo {author}
  {\bibfnamefont{Y.-Q.}\ \bibnamefont{Wang}}, \bibinfo {author}
  {\bibfnamefont{B.}~\bibnamefont{Lorenz}},\ and\ \bibinfo {author}
  {\bibfnamefont{C.~W.}\ \bibnamefont{Chu}},\ }%
  \bibfield{journal}{%
  \bibinfo {journal} {Phys. Rev. B}\ }%
  \textbf{\bibinfo {volume} {83}},\ \bibinfo {pages} {140401(R)} (\bibinfo
  {year} {2011})%
  \bibAnnoteFile{NoStop}{ye:11}%
\bibitem{meier:09}%
  \BibitemOpen
  \bibfield{author}{%
  \bibinfo {author} {\bibfnamefont{D.}~\bibnamefont{Meier}}, \bibinfo {author}
  {\bibfnamefont{M.}~\bibnamefont{Maringer}}, \bibinfo {author}
  {\bibfnamefont{T.}~\bibnamefont{Lottermoser}}, \bibinfo {author}
  {\bibfnamefont{P.}~\bibnamefont{Becker}}, \bibinfo {author}
  {\bibfnamefont{L.}~\bibnamefont{Bohaty}},\ and\ \bibinfo {author}
  {\bibfnamefont{M.}~\bibnamefont{Fiebig}},\ }%
  \bibfield{journal}{%
  \bibinfo {journal} {Phys. Rev. Lett.}\ }%
  \textbf{\bibinfo {volume} {102}},\ \bibinfo {pages} {107202} (\bibinfo {year}
  {2009})%
  \bibAnnoteFile{NoStop}{meier:09}%
\end{thebibliography}

%

\end{document}